\documentclass[12pt,preprint]{aastex}

\shorttitle{FUV Flux in Young Embedded Star Clusters}
\shortauthors{Holden et al.}

\newcommand{\be}{\begin{equation}}
\newcommand{\ee}{\end{equation}} 

\begin{document}

\title{An Investigation of the Loss of Planet-Forming Potential in\\
    Intermediate Sized Young Embedded Star Clusters}

\author{Lisa Holden, Edward Landis, and Jeremy Spitzig}
\affil{Department of Mathematics and Statistics, Northern Kentucky University,
    Highland Heights, KY 41099}   

\and
\author{Fred C. Adams}
\affil{Department of Physics, University of Michigan, Ann Arbor, MI 48109}   

\begin{abstract}

A large fraction of stars forming in our galaxy are born within
clusters embedded in giant molecular clouds.  In these environments,
the background UV radiation fields impinging upon circumstellar disks
can often dominate over the radiation fields produced by each disk's
central star.  As a result, this background radiation can drive the
evaporation of circumstellar disks and lead to the loss of planet
forming potential within a cluster.  This paper presents a detailed
analysis of this process for clusters whose stellar membership falls
within the range $100 \le N \le 1000$.  For these intermediate-sized
clusters, the background UV field is often dominated by the most
massive stellar member.  Due to the steep slope of the initial mass
function, the amount of background UV light that bathes clusters of
similar size displays significant variance. As a result, we perform a
statistical analysis of this problem by calculating distributions of
FUV flux values impinging upon star/disk systems for several cluster
scenarios.  We find that in the absence of dust attenuation, giant
planet formation would likely be inhibited in approximately half of
systems forming within intermediate-sized clusters regardless of
stellar membership.  In contrast, the presence of dust can
significantly lower this value, with the effect considerably more
pronounced in more populated clusters.

\end{abstract}


\section{Introduction}

The formation of stars within the Milky Way is presently occurring
within massive complexes of molecular gas and dust known as giant
molecular clouds (GMCs).  These highly non-uniform structures
typically contain several dense clumps with characteristic densities
$n \sim 10^3$ cm$^{-3}$ and radii in the range $R = 0.2 - 2$ pc. The
largest of these clumps contain as many as $\sim 1000$ small ($R \sim
0.1 - 0.2$ pc), dense ($\sim 10^4 - 10^5$ cm$^{-3}$) condensations
called cores, which are the sites of individual star formation events.
The mass function of these cores has been measured to span the mass
range $\sim 1 - 100 M_\odot$ with a peak $\sim 10 M_\odot$ (Jijina et
al. 1999), and, more recently, a mass range $\sim 0.2 - 20 M_\odot$
and a characteristic mass of $\sim 2 M_\odot$ (Lada et al. 2008). On
the theoretical front, competing paradigms have been developed over
the past two decades to explain how the dense regions of GMCs become
gravitationally unstable (e.g., Shu et al. 1987; Myers 1998; Klessen
\& Burkett 2000, 2001; McKee \& Ostriker 2007), and subsequently, how
the ensuing collapse forms a protostar/disk system (e.g., Shu 1977;
Fatuzzo et al. 2004).

Although many of the details describing how stars form remain under
study, it is clear that the end result of this complex process is the
formation of stellar nurseries containing tens to thousands of members
(e.g., Lada \& Lada 2003; Porras et al. 2003).  Indeed, young embedded
clusters appear to be basic units of star formation (e.g., Gutermuth
et al. 2009), and they account for perhaps 90\% of the stars that
populate the Galactic disk.  A significant body of work now exists on
how these young clusters evolve and how their environment affects the
formation processes occurring within. Not surprisingly, early work 
focused on either small ($N \le 100$) clusters (e.g., Lada et al. 1984)
or large ($N \ge 10,000$) clusters (e.g., Portegies Zwart et al. 1998;
St\"orzer \& Hollenbach 1999; Boily \& Kroupa 2003).  However, about
60\% of stars observed in nearby ($\la 2$ kpc) embedded clusters
belong to intermediate-sized systems with $N = 100 - 1000$ (Lada \&
Lada 2003; Porras et al. 2003; see also Adams \& Myers 2001).

Motivated by these observational results, Adams et al. (2006)
performed a suite of $N$-body simulations to explore how the evolution
of intermediate sized clusters depends on the system size $N$ and on
initial conditions.  One aspect of this previous work was an analysis
of how the background FUV fields in embedded clusters affect planetary
formation.  Toward that end, these authors calculated a probability
distribution for the FUV flux experienced by the ensemble of cluster
stars in the Lada \& Lada (2003) catalog as a function of the FUV
flux.  Specifically, for each star within a cluster of size $N$, a
Salpeter IMF was sampled $N$ times to determine a corresponding
cluster background FUV luminosity, where the models of Maeder \&
Meynet (1987) and Schaller et al. (1992) were used to specify the FUV
luminosity as a function of stellar mass.  A stellar density profile
of the form $\rho_* \propto r^{-1}$ for $0\le r \le R_{c}$ was then
sampled to calculate an effective flux exposure, where the cluster
size $R_{c}$ was based on a fit to the data obtained by Lada \& Lada
(2003) and Carpenter (2000).  This process was then repeated for all
clusters within the catalog in order to ascertain what fraction of
stars have orbit-averaged fluxes exceeding benchmark values for which
the effects of photoevaporation on circumstellar disks have been
calculated in detail (Adams et al. 2004).  It is important to note
that the effects of dust attenuation were not included in these
calculations.  Since a substantial amount of gas and dust can be
present in the earliest stages of cluster evolution, these results
provide an upper limit to the effects of radiation on planetary
formation.

The aforementioned statistical investigation was extended by
considering how much radiation a given solar system experiences over
the course of its orbit.  Specifically, the authors used approximate
analytical descriptions of orbits in a Hernquist potential (Hernquist
1990) in order to obtain a simple expression (Adams \& Bloch 2005) for
the mean flux experienced by a star in a cluster of size $N$ in terms
of the star's (dimensionless) energy $\epsilon$ and angular momentum
$q$ (see equation [\ref{eqdefs}] for definitions).  This expression
was later incorporated in follow-up work on the effects of both EUV
and FUV background fields on circumstellar disks and planetary
formation (Fatuzzo \& Adams 2008; see also Armitage 2000).
Specifically, the distributions of orbit-averaged fluxes for both EUV
and FUV radiation bands were constructed by sampling the orbital
parameters $\epsilon$ and $q$.  In addition, these authors also
considered the effects of dust attenuation in their analysis, and
thereby also provided limits on the orbit-averaged flux impinging on
young star/disk systems.

This paper builds upon previous work in three important ways: First,
this treatment presents a considerably more detailed analysis of how
stellar orbits affect the amount of radiation impinging upon the
surrounding protoplanetary disks; this work includes generalized forms
for the cluster potentials.  Second, the stellar orbits are calculated
numerically rather than through the previously adopted analytic
approximations, thereby yielding greater accuracy in the desired
output measures. [For completeness, we also note that the analytical
expression presented in Adams et. al (2006) contained a typo so that
the results of Fatuzzo \& Adams (2008), which used this expression,
correspond to the case of nearly circular orbits.]  Finally, several
different cluster environment scenarios are considered in detail,
thereby allowing us to explore how cluster size $N$, mass density
profile, star formation efficiency, and dust attenuation affects
planetary formation.

The timescales of interest in this paper --- for cluster evolution,
disk evolution, and planet formation --- all lie in the range 1 -- 10
Myr. Young embedded clusters typically retain their gaseous component
for 3 -- 5 Myr (Gutermuth et al. 2009) and remain intact for $\sim10$
Myr (Lada \& Lada 2003; Porras et al. 2003).  The lifetime of
circumstellar disks is comparable; more specifically, the fraction of
the observed disk population that retains signatures of nebular gas is
a decreasing function of time, with a half-life of about $\sim3$ Myr
and an e-folding time of $\sim 5$ Myr (Hern{\'a}ndez et al. 2007). On
a related note, the time required for typical radiation fields to
evaporate the outer portions of circumstellar disks is also measured
in Myr (Johnstone et al. 1998; Adams et al. 2004; Ercolano et al.
2009).  Finally, the timescale for giant planet formation through the
core accretion mechanism (Lissauer \& Stevenson 2007) is typically 2
-- 7 Myr, while the timescale for planet migration is somewhat shorter
at $\sim1$ Myr (Papaloizou \& Terquem 2006).

Although this paper focuses on the effects of radiation, which leads
to mass loss in circumstellar disks, we note that cluster environments
provide additional influences on forming solar systems: Dynamical
interactions between the solar systems and other cluster members can
produce tidal truncation of the disks (Clarke \& Pringle 1993;
Kobyashi \& Ida 2001) and/or disruption of the planetary orbits at
later times (Adams et al. 2006; Malmberg \& Davies 2009). On a related
note, the disks can gain mass through Bondi-Hoyle accretion as they
orbit through the cluster (Throop \& Bally 2008).  In cluster
environments, massive stars can also influence solar systems through
supernova explosions, which can sculpt disks through interactions with
their blast waves (Chevalier 2000; Ouelette et al. 2007) and by
enriching the disk material with radioactive material (Cameron \&
Truran 1977; Williams \& Gaidos 2007). 

The paper is organized as follows.  We review the physical
characteristics of intermediate sized clusters in \S 2, with a focus
on cases with $N$ = 100, 300, and 1000.  We then consider the orbits
of stars that populate these clusters in \S 3 for two different static
density profiles assumed as representative of the true stellar/gas
content of the system.  In \S 4, we compute the orbit-averaged FUV
flux over a range of orbital parameters for several cluster profiles,
obtaining values for cases with and without dust extinction. A
statistical analysis is performed in \S 5 in order to quantify the
expected effect on disk-evaporation in clusters of membership size $N$
= 100, 300, and 1000. The paper then concludes, in \S 6, with a
summary of results and a discussion of their implications.

\section{The Stellar Populations of Intermediate Sized Clusters}

Embedded cluster environments, and in particular, the radiation fields
which bathe them, depend sensitively on the population of their
stellar members.  We consider here the properties of intermediate
sized clusters with membership numbers within the range $100 \leq N
\leq 1000$.  Of course, such a characterization can only be done
statistically, as two clusters with the same number of members could
have considerably different populations, owing in large part, to the
steepness of the initial mass function (IMF). In addition, the (few)
large stars produce the most radiation.  As a result, most of the FUV
light bathing intermediate-sized clusters originates from the most
massive stellar member.  Given the steepness of the initial mass
function, and hence the rarity of massive stars, there is therefore a
large variance in the total FUV luminosities of intermediate-sized
clusters.

As a starting point, we begin by considering the distribution of
star-forming cluster sizes, defined in terms of stellar membership
$N$.  Recent infrared observations have produced catalogs of embedded
clusters out to 1 kpc (Porras et al. 2003) and 2 kpc (Lada \& Lada
2003).  Of course, both catalogs are almost certainly not complete,
with stellar memberships of larger clusters undercounted due to the
inability of fully observing all of the faint, low-mass members.  In
addition, it is also possible that smaller clusters ($N \la 100$) have
been missed altogether.  Nevertheless, these catalogs are likely
representative of the population of intermediate clusters within the
disk of our galaxy.  The fraction $f(N)$ of the total number of stars
in the union of these catalogs is shown in Figure 1.  As can be seen
by the dotted reference lines in this figure, nearly 60\% of these
stars are born in clusters with $100 \leq N \leq 1000$, and are evenly
distributed logarithmically in $N$ within this range (shown by the
linear fit to the data within this range as denoted by the solid
curve).  The characteristic membership size for these intermediate
clusters is thus $N_{ch} \approx 300$, as illustrated by the dashed
reference line in Figure 1.

Young embedded clusters are observed to have radii $R_{c}$ 
ranging between $0.1 - 2$ pc, with a clear correlation between 
$R_c$ and $N$, although significant scatter exists in the data
(Carpenter 2000; Lada \& Lada 2003).  We use here the relation
\begin{equation}
R_c = 1.0\, {\rm pc}\,\sqrt{(N/300)}\;,
\label{radlaw} 
\end{equation}
which is obtained through a simple fit to the aforementioned data (see
Figure 1 of Adams et al. 2006). Given that $\sim 90$\% of stars have
masses less than $1 M_\odot$, the total stellar mass of a cluster of
size $N$ is expected to be $M_{*} \approx 0.5 N\, M_\odot$.  We assume
here that this stellar content is contained within $R_{c}$.  Of
course, the gas from which these stars are born extends much farther,
eventually merging smoothly into the GMC background. This gaseous
component is eventually disrupted through the action of stellar winds
and outflows, radiative processes, and supernovae (e.g., Whitworth
1979; Matzner \& McKee 2000; Gutermuth et al. 2004).  Although the
details of these processes are not fully understood, observations
indicate that clusters older than 5 Myrs are rarely associated with
molecular gas, so that gas removal must occur on a timescale of 
$\sim 3 - 5$ Myr (Lada \& Lada 2003).

For an IMF characterized by a probability function $dP/dm$, 
the expectation value $\langle L_{FUV}^* \rangle$ of the
FUV luminosity {\it per star} is determined by the integral 
\begin{equation}
\langle L_{FUV}^* \rangle = \int^{m_{max}}_{m_{min}} 
L_{FUV}^* (m) {dP\over dm} \;dm\,,
\label{lexpect} 
\end{equation} 
where $L_{FUV}^* (m)$ is the stellar FUV luminosity as a function of
mass $m$ (in units of $M_\odot$). We note that the integral is not
sensitive to any reasonable value of minimum mass $m_{min}$, because
most of the FUV light is produced by the most massive stars, in spite
of their smaller numbers.  Throughout this work, we adopt a stellar
IMF with a Salpeter slope at high masses and an upper cutoff $m_{max}$
= 100. With this IMF, the integral of equation (\ref{lexpect}) yields
a value of $\langle L_{FUV}^* \rangle = 1.23 \times 10^{36}$ ergs
s$^{-1}$.  In principle, the value $N \langle L_{FUV}^* \rangle$ can
be used to characterize the ``typical'' FUV luminosity of a cluster of
size $N$.  However, because of the steepness of the IMF, only a small
number of high-mass stars are likely to populate intermediate-sized
clusters.  As a result, the distributions of the maximum stellar
masses and FUV luminosities for a population of clusters of size $N$ are
heavily skewed, so that the ``typical'' maximum stellar mass and FUV
luminosity of a cluster are better represented by the median values of
the distributions (which are significantly smaller than the mean
values).  In addition, the total cluster luminosity $L_{FUV}^c$ is
dominated by the the luminosity $L_{FUV}^{max}$ of the most massive
star.

To illustrate these points, we perform a suite of simulations in order
to build up distributions of $m_{max}$, $L_{FUV}^{max}$ and
$L_{FUV}^c$ for clusters with sizes $N$ = 100, 300, and 1000.
Specifically, we sample our adopted IMF $N$ times in order to get the
masses of the $N$ stars which populate our simulated cluster.  We use
the results presented in Armitage (2000) to obtain the FUV luminosity
of each stellar member, thereby finding the total cluster FUV
luminosity.  We also find the FUV luminosity of the most massive star
in the cluster.  This procedure is then repeated 100,000 times to
build a distribution of luminosities (normalized by $\langle L_{FUV}^*
\rangle$) and maximum stellar mass (in units of $M_\odot$) for each
assumed cluster size $N$.  

The luminosity distributions are shown in Figures 2 -- 4 for $N$ =
100, 300, and 1000, respectively, where the unshaded (shaded) portions
represent the cluster (most massive star) luminosities.  The vertical
solid and dashed lines denote the median values of the cluster and of
the most massive star luminosities, respectively.  In addition, one
sees that for $N \la 300$, the most massive stars do indeed dominate
the FUV luminosity of their cluster.  For $N = 1000$, the most massive
star typically provides about half of the total cluster luminosity.
The mean and median values of our obtained distributions are
summarized in Table 1. As expected, the median values fall well below
the expectation values.
 
\section{Orbit Solutions}

This work focuses on how the average flux that impinges upon a
circumstellar disk depends upon the orbit of its parent star and the
cluster environment in which it resides.  While encounters with other
stellar members can alter a star's orbit, the resulting effects for
intermediate sized clusters are relatively small (Adams et al. 2006)
over the timescales for disk evolution and planet formation (3 -- 10
Myr).  As a result, one can obtain good approximations for stellar
orbits by assuming that the stars move through a smooth and
time-independent gravitational potential that includes both the gas
and stellar content of the cluster.  To model this potential, we
consider two density profiles of the form 
\begin{equation}
\rho(\xi) = \frac{\rho_{0}}{\xi(1+\xi)^{c}}, 
\end{equation}
where $\xi = r/r_s$ is a dimensionless radius and $r_s$ is a
length scale that characterizes the width of the density profile 
(the radial size of the cluster). Here we consider two values for 
the index: $c$ = 3, 4 (see below). This assumed form describes
observed clusters in that the density scales as $1/\xi$ for small
values of $\xi$ and drops off quickly into the background for large
values of $\xi$. Note that the radial dependence $\rho \sim 1/\xi \sim
1/r$ for gas in clusters (and cluster-forming cores) has been inferred
from observations (e.g., Larson 1985; Jijina et al. 1999). For
simplicity, we set $r_s = R_c = 1.0$ pc $\sqrt{N/300}$ throughout this
paper (consistent with the observations of Lada \& Lada [2003] and the
scaling law of equation [\ref{radlaw}]).  As shown below, orbit
solutions can be obtained in terms of two dimensionless quantities
that characterize the energy and angular momentum of the stellar
orbit. These orbits can then be placed in a desired cluster
environment through the proper choice of the density scale $\rho_0$,
as carried out in \S 4.

We characterize the orbit solutions for both the Hernquist potential
($c=3$) and the steeper modified Hernquist potential ($c=4$). Both
cases yield analytical results and thereby provide a convenient
framework from which to explore the orbit solutions. The orbit
solutions for the Hernquist potential have been explored in detail
(Adams \& Bloch 2005). Here we summarize these previous results and
extend the analysis to include the modified Hernquist potential. The
Hernquist gravitational potential ($c=3$) and corresponding mass
profile are given by the expressions
\begin{equation}
\Psi (\xi) = \frac{\Psi_{0}}{1+ \xi},  \quad\quad\quad 
M(\xi) = M_{\infty}\frac{\xi ^{2}}{(1+\xi)^{2}}\,,
\end{equation}
where $\Psi_{0} \equiv 2 \pi G \rho_{0} r_{s}^{2}$ is the total depth
of the gravitational potential and $M_{\infty} \equiv 2 \pi r_{s}^{3}
\rho_{0}$ is the total mass.  The corresponding expressions for the
modified Hernquist gravitational potential ($c=4$) are
\begin{equation}
\Psi (\xi) = \Psi_{0} \frac{ (2+\xi)}{2(1+ \xi)^2},  
\quad\quad\quad M(\xi) = M_{\infty}\frac{\xi^3+3\xi ^{2}}{(1+\xi)^{3}}\,,
\end{equation}
where $\Psi_{0} \equiv 4 \pi G \rho_{0} r_{s}^{2}/3$ and $M_{\infty}
\equiv 2 \pi r_{s}^{3} \rho_{0}/3$.  Note that all quantities are
taken to be positive, with the proper signs inserted as necessary.

\subsection{General Orbits} 

Orbits in spherical mass distributions are governed by the following
differential equation (Binney \& Tremaine 1987)
\begin{equation}
{d\theta \over dr} = {1\over r}\left[ 
{2(E-V)r^{2}\over j^{2}}-1 \right] ^{-1/2},
\label{orbbasic} 
\end{equation}
where $V$ is the potential, $E$ is the energy, and $j$ is the specific
angular momentum. We consider only bound orbits with negative
energy. Following Adams \& Bloch (2005), we introduce the
dimensionless variables
\begin{equation}
\epsilon \equiv \left | E \right |/\Psi_{0}, \quad\quad\quad  
q \equiv j^{2}/2 \Psi_{0} r_{s}^{2}.
\label{eqdefs} 
\end{equation}
Equation (\ref{orbbasic}) can thus be written in dimensionless form as
\begin{equation}
\xi {d\theta \over d\xi} = [q/f(\xi)]^{1/2},
\end{equation}
where the rational function $f(\xi)$ determines the properties of the
orbits. For the Hernquist potential, this function takes the form
\begin{equation}
f_{H}(\xi) = -\epsilon \xi ^{2}+{\xi ^{2} \over 1+\xi}-q \, ,
\end{equation}
whereas the modified Hernquist potential yields the form 
\begin{equation}
f_{M}(\xi) = -\epsilon \xi ^{2} + 
{\xi^2(2+\xi) \over 2 (1+\xi)^{2}} - q \, .
\end{equation}

Orbits can exist only over radii where the function $f(\xi)$ is
positive. For values of the parameters $\epsilon$ and $q$ that
correspond to bound orbits, the expression $f(\xi)$ has two real
positive zeroes, $\xi_{1}$ and $\xi_{2}$, such that $f(\xi)$ is
positive for $\xi_{1} < \xi < \xi_{2}$.  The values $\xi_{1}$ and
$\xi_{2}$ correspond to the radial turning points of the orbit, as
shown by the spirographic orbit depicted in Figure 5. For a given
value of $c$, the turning points may need to be found numerically for
each set of $\epsilon$ and $q$.  However, the inverse relations can
be written down explicitly (analytically) and take the forms
\begin{equation}
\epsilon = \frac{ \xi_{1} (1+\xi_{2})^{c-2} - \xi_{2} (1+\xi_{1})^{c-2} 
+ (\xi_{2} - \xi_{1})(1+\xi_2)^{c-2}(1+\xi_{1})^{c-2}}   
{(\xi_{2}-\xi_{1})(\xi _{1}+\xi _{2})
(c-2)(1+\xi_{1})^{c-2}(1+\xi_{2})^{c-2}  } ,
\end{equation}
\begin{equation}
q=\xi_{1}\xi_{2} \frac{ \xi_{1} (1+\xi_{1})^{c-2} - \xi_{2} (1+\xi_{2})^{c-2} 
+ (\xi_{2} - \xi_{1})(1+\xi_2)^{c-2}(1+\xi_{1})^{c-2}}   
{(\xi_{2}-\xi_{1})(\xi_{1} + \xi_{2})(c-2)(1+\xi_{1})^{c-2}
(1+\xi_{2})^{c-2}  } \, . 
\end{equation} 
For the Hernquist profile ($c=3$), these expressions reduce to
previously known results [see equation (6) from Adams \& Bloch
(2005)]; for the modified Hernquist potential ($c=4$) these results 
are new. Notice also that the term $(\xi_{2}-\xi_{1})$ term can be
factored out of the numerator, for both expressions and for both $c$ =
3 and 4; these expressions are thus non-singular.

In contrast to orbits in a Keplerian potential, the energy 
parameter in cluster potentials varies over a finite range, where 
0 $< \epsilon <$ 1. The value of $\epsilon$ determines the radial
scale of the orbit, where $\epsilon = 0$ corresponds to a star that is
outside the potential well of the cluster environment ($r \rightarrow
\infty$) and $\epsilon = 1$ corresponds to a star at rest in the
cluster center ($r \rightarrow 0$). The angular momentum parameter $q$
characterizes the effective ``eccentricity'' of the spirographic
orbit, with $q = 0$ corresponding to a purely radial orbit.  For a
specific value of the energy parameter $\epsilon$, the maximum value
of $q$ for which bound orbits exist occurs where $\xi_{1}$ and
$\xi_{2}$ coincide, and this case thus corresponds to a circular
orbit. For the Hernquist potential, the maximum value $q_{max}$ in
terms of $\epsilon$ can be found analytically by solving equations
(11) and (12) with $\xi_1 = \xi_2$, yielding the expression
\begin{equation}
q_{max} = {1\over \epsilon} 
{(1+\sqrt{1+ 8\epsilon} - 4\epsilon)^3 \over (1+\sqrt{1+ 8\epsilon})^2}\,.
\end{equation}
For the modified Hernquist potential, this procedure requires the
solution of a cubic polynomial and is most easily performed
numerically.

\section{Orbit Average Flux Calculations}

The FUV flux impinging upon the circumstellar disk surrounding a star
belonging to a young cluster can vary greatly over the course of the 
stellar orbit.  Taking into account this effect, we can gauge the
efficiency with which the background FUV luminosity can drive the
evaporation of circumstellar disks by calculating orbit-averaged FUV
flux values for several different cluster profiles, as defined by the
four following criteria:

{\sl Cluster membership N.}  -- As shown in Figure 1, the majority of
stars appear to form in clusters with stellar membership $N$ in the
range $100 \le N \le 1000$.  We sample this range by considering three
values of cluster membership: $N = 100$, $300$ and $1000$.  For each
choice of $N$, the cluster radius is set through the relation given by
Equation (1), and the total stellar mass for the cluster is taken to
be $M_* = 0.5 N M_\odot$.  For simplicity, we consider only the FUV
luminosity produced by the most massive stellar member, which is
assumed to reside at the center of the cluster.  Given the skewed
nature of the distributions of cluster luminosities obtained in \S 2,
we adopt the median values listed in Table 1 for both the luminosity
and maximum stellar mass.

{\sl Star formation efficiency.} -- Observational determinations of the
star formation efficiency in nearby embedded clusters, though
uncertain, range between $\eta = 0.1 - 0.3$ (Lada \& Lada 2003).  We
consider two values -- the midpoint value $\eta = 1/5$ of this range
and the value $\eta = 1/3$ as adopted by Adams et al. (2006).  We
assume that all stars form within a radius $R_c$, so that $\eta =
M_*/M_1$ where $M_1$ is the {\it total} mass of the gas and stars
contained within the radius $R_c$.

{\sl Gas density profile.} -- We assume that the gas and stellar
contents of a young cluster are suitably described by a smooth mass
density profile of the form given in Equation (3) with $r_s = R_c$,
and consider cases for which $c = 3$ (Hernquist profile) and $c = 4$
(modified Hernquist profile).  We connect the value of $\rho_0$ to
each scenario through the relations
\begin{equation}
\rho_0 = {2 M_1\over \pi r_s^3} \qquad ({\rm for} \, \, c = 3)\, , 
\qquad {\rm and} \qquad \rho_0 = {3 M_1\over \pi r_s^3} \qquad 
({\rm for} \, \, c = 4) \, , 
\end{equation}
where $M_1=\int_0^{R_c} {4\pi r^2\rho(r)dr} = M_\infty/4$ if $c=3$ and
$M_1=M_\infty/2$ for $c=4$.  Expressing $\rho_0$ in terms of the star
formation efficiency and stellar membership, we find that
\begin{equation}
\rho_0 = {300 M_\odot \over \pi (1.0  \,{\rm pc})^3}\, \eta^{-1}
\left({N\over 300}\right)^{-1/2} \quad\quad\quad 
({\rm for} \quad c=3) \, ,
\end{equation}
and 
\begin{equation}
\rho_0 = {900 M_\odot \over 2\pi (1.0  \,{\rm pc})^3}\, \eta^{-1}
\left({N\over 300}\right)^{-1/2} \quad\quad\quad 
({\rm for} \quad c=4) \, .
\end{equation}

{\sl Dust Attenuation.} -- We calculate the orbit-averaged flux both
with and without dust attenuation.  When considering dust attenuation,
we assume that a mass $Q m_{max}$ of gas and dust collapses to form
the central star, and consider values of $Q = 3$ and $Q = 10$ in our
analysis. Since $Q > 1$, we are thus assuming that the star formation
process in the central region of the cluster is not 100 percent
efficient.  The radius of the void surrounding the central star is
obtained by setting $Q m_{max}$ equal to $M(\xi_c)$, thereby yielding
\begin{equation}
\xi_c = \frac { Qm_{max} + \sqrt{Qm_{max} 2 \pi\rho_0 R_c^3  }} 
{ 2\pi\rho_0R_c^3 - Qm_{max}  } \;
\end{equation}
when $c = 3$.  As with $q_{max}$, this quantity is best obtained
numerically when $c = 4$.  We assume a gas density $\rho_g (\xi) =
(1-\eta)\rho(\xi)$ beyond a radius $\xi_c$.  The optical depth outside
the void is therefore given by the expression
\begin{equation}
\tau_{FUV} (\xi) = N_{col} (\xi)\, \sigma_{FUV} = 
R_c \,\sigma_{FUV} \int_{\xi_c}^\xi {{\rho_g(\xi')\over m_H}} d\xi'\,,
\end{equation}
where we adopt a value of $\sigma_{FUV} = 8\times 10^{-22}$ cm$^2$ for
the dust cross-section per Hydrogen nucleus (St\"orzer \& Hollenbach
1999).  We note that the integral in equation (18) can be solved
analytically for both $c = 3$ and $c = 4$ .
 
For convenience, we introduce a dimensionless time $\tau \equiv t /\tau_{0}$, 
where $\tau_0 \equiv r_s/\sqrt{2\Psi_0}$.  The (dimensionless) orbital period 
can be obtained through the expression
\begin{equation}
\tau_{orb} = 2 \int _{\xi_{1}}^{\xi_{2}} 
\frac {\xi} { \sqrt{f(\xi)}} d\xi \, , 
\end{equation}
and is shown in Figure 6 as a function of dimensionless energy
$\epsilon$ for our three chosen values of $N$ and the two density
profiles (we note that $\tau_{orbit}$ is not sensitive to $q$). The
corresponding values of the orbital period are smaller for the $c = 4$
case, as expected, given the proportionality $\tau_0 \propto
\rho_0^{-1/2}$.  These results indicate that stars will complete at
least one orbit before the gas and dust are removed by the action of
stellar winds (which, as noted in \S 2, occurs on a timescale of $3 -
5$ Myr).

In the absence of dust attenuation, the orbit-averaged FUV flux is
given by the expression
\begin{equation}
\langle F_{FUV}\rangle  =  {L_{FUV}^{max} \over 4 \pi R_c^2}
{1 \over \tau_{orb}}
\int_0^{\tau_{orb}} {d \tau \over \xi^2}\;,
\label{meanflux} 
\end{equation}
where the orbital radius $\xi$ of the star as a function of time is
obtained by numerical integration of the governing force equations.
The effects of dust attenuation are easily included by multiplying the
integrand in equation (\ref{meanflux}) by a factor of
$\exp[-\tau_{FUV}]$, where $\tau_{FUV}$ is the optical depth to FUV
radiation for a given point along the orbit. 

We calculate values of the orbit-averaged flux for a range of energies
$\epsilon$ corresponding to orbits with radii less than $R_c$ and at three
values of $q/q_{max}$ for 27 cluster profiles (defined through a
choice of $N$, $\eta$, $c$ and dust scenario).  A subset of these
results are presented in Figures 7 -- 9.  In each figure, the
horizontal dotted line represents our benchmark value for determining
whether a the planet-forming potential of a circumstellar disk has
been compromised by the background FUV radiation. As a working
benchmark value, we use the flux $G_B = 4.8$ ergs s$^{-1}$ cm$^{-2}$ =
3000 $G_0$ (where $G_0 = 1.6\times 10^{-3}$ ergs s$^{-1}$ cm$^{-2}$);
with this radiation level, the disk surrounding a star with mass
$M_\ast$ = 1 $M_\odot$ would be reduced to a radius of $\sim36$ AU
over the course of 10 Myr (Adams et al. 2004).  Since giant planets
form in the 5 -- 30 AU region, over a comparable timescale, star/disk
systems that experience an average flux greater than $\sim G_B$ are
expected to have a diminished planet-forming potential.  The
truncation of circumstellar disks exposed to a given level of FUV flux
depends on the mass of the central star. For the benchmark flux value
$G_B$, a star with mass $M_\ast$ = 0.5 (0.25) $M_\odot$ will evaporate
down to 18 (9) AU over the course of 10 Myr.

The full set of results for all of the sampled cluster environments
are qualitatively described by Figures 7 -- 9.  In addition, we can
summarize these results by specifying the value of energy $\epsilon_b$
for which the orbit-averaged flux equals our benchmark value. These
values of $\epsilon_b$ are listed in Table 2.

\section{Statistical Analysis for Clusters}

As illustrated in Figures 2 -- 4, the steep nature of the stellar IMF
results in a large variance in the FUV luminosities for intermediate
sized clusters with the same number of stars.  Subsequently, the
efficiency with which circumstellar disks are evaporated by background
FUV fields can vary greatly from cluster to cluster.  We therefore
present a statistical analysis of the amount of FUV radiation
impinging upon cluster members. Since one focus of this work is to
determine how the cluster membership size $N$ affects planetary
formation, we first adopt values of $c = 3$ and $\eta = 1/5$ (as
suggested from observations), and calculate the distribution of FUV
flux values experienced by stars for clusters of size $N$ = 100, 300,
and 1000.  We note that the evaporation of stars would be more
pronounced for a higher star formation efficiency and a steeper
density profile, as can be seen by comparing values of $\epsilon_b$
for the various cases presented in Table 2.  Flux values are
calculated for both the case of no dust attenuation and the case with
dust attenuation and mass depletion parameter $Q = 3$. These two
classes of results thus represent upper limits (no dust) and lower
limits (maximum dust) to the average flux experienced by the stars
within the specified cluster environment.

The stellar dynamics of a cluster determines the distribution of
energy and angular momentum for the stellar members (e.g., Binney \&
Tremaine 1987). One can therefore find the relationship between the
differential energy distribution, and the density profile of the
cluster for a given set of assumptions about the velocity
distribution.  We adopt here an isotropic velocity distribution and a
density profile form $\rho \propto 1/r$, corresponding to the limit
regimes of our chosen density profiles (since we assume that stars are
formed within a radius $R_c$).  For this case, the differential energy
distribution --- the probability distribution for orbital energies ---
takes the form
\begin{equation}
h(\epsilon) = {d P_m \over d \epsilon} = 
{2 \over (1 - \epsilon_0)^2 } (1 - \epsilon) \, , 
\end{equation}
which is normalized for the range of dimensionless energies
$\epsilon_0 \le \epsilon \le 1$. Given our assumption that stars
are found within $R_c$,  we truncate the distribution at
$\epsilon_0 = 0.375$. 

The probability distribution for orbital angular momentum depends on
the initial velocities with which the stars are born. Recent
observations of pre-stellar cores suggest that these starting
velocities are lower than the values expected for virial equilibrium
(e.g., Andr\'e 2002; Peretto et al. 2006). One might therefore expect
that stars in clusters have preferentially lower values of
$q/q_{max}$.  However, these lower initial velocities allow the entire
cluster to shrink after the stars form and enter into ballistic
orbits.  In the resulting smaller dynamical system, smaller amounts of
angular momentum ``go farther'' and result in somewhat larger $q$
values for stellar orbits in the cluster. As a result, stars in
nascent clusters tend to have intermediate values of $q$ (Adams et
al. 2006).  For simplicity we assume that the probability distribution
for orbital angular momentum is uniform (though we consider a
particular case with circular orbits below).

For each cluster size, the adopted IMF (see \S 2) is sampled $N$ times
in order to obtain a value for the most massive star in the cluster,
which is assumed to be centrally located.  The results presented in
Armitage (2000) are then used to obtain a corresponding FUV luminosity
(see also Maeder \& Meynet 1987; Schaller et al. 1992).  Values of
$\epsilon$ and $q/q_{max}$ are randomly selected from their
corresponding probability distributions, and the procedure outlined in
\S 4 is then used to calculate the orbit-averaged flux for the cases
of no dust attenuation and dust attenuation with $Q = 3$.  This
process is repeated 10,000 times in order to build up a sample
distribution.  The results are presented in Figures 10 -- 12 for the
cases of $N$ = 100, 300, and 1000, respectively, where the unshaded
histogram represents the case of no dust attenuation and the shaded
histogram represents the case with dust attenuation.  The solid
vertical line denotes our benchmark value $G_B = 3000\,G_0$, and the
vertical dashed lines denote the median values of the distributions
(the lower value corresponding to the case with dust attenuation).

As illustrated by the median values of the calculated distributions,
the radial size of a cluster does not significantly affect the amount
of FUV flux impinging upon its members in the absence of dust
attenuation (this result follows from the assumed/observed scaling law
of equation [\ref{radlaw}] for cluster radii).  To quantify this point
further, we note that 37\%, 53\%, and 63\% of the stars in the
populations shown by the unshaded histograms (no dust) in Figures 10,
11 and 12, respectively, experienced FUV flux values greater than our
benchmark value. In contrast, the presence of dust can significantly
reduce the FUV flux impinging upon cluster members, with larger
clusters better able to shield stellar systems from the central FUV
radiation. Interestingly, however, 30\%, 33\%, and 27\% of the stars
in the populations shown by the shaded histograms (dust) in Figures
10, 11 and 12, respectively, experienced FUV flux values greater than
our benchmark value. Thus, although stars are better shielded from the
FUV radiation in larger clusters on the whole, the efficiency with
which the formation of giant planets is inhibited (in the presence of
dust attenuation) is nearly the same in all medium-sized clusters.
This universality results from the fact that while larger clusters
have greater column densities, the net effect of dust attenuation is 
offset by the larger central void regions associated with more massive
stars.  As such, stars in smaller clusters are as likely to enter the
central void as their counterparts in larger clusters. Indeed, the
high-flux peaks of the distributions shown in Figures 10 -- 12 for the
case of dust attenuation represent the stars that enter this void.

The results presented in \S 4 suggest that disk photoevaporation is
more efficient for clusters with steeper density profiles. This trend
occurs because clusters with steeper density profiles have a greater
concentration of mass at their centers; as a result, orbits with a
given energy $\epsilon$ will be tighter, leading to an increase in
radiative flux. To illustrate this point further, we calculate the
distribution of FUV flux values experienced by stars for a cluster of
size $N = 300$ and a star formation efficiency of $\eta = 1/5$ (as
with the case presented in Figure 11), but with a steeper $c = 4$
density profile. The result is presented in Figure 13. Comparing with
Figure 11, we find that the median flux values --- both for the case
of no dust and dust attenuation --- are greater for the $c$ = 4 case.
Indeed, 68\% of the stars in the population shown by the unshaded
histogram (no dust) in Figure 13 experienced FUV flux values greater
than our benchmark value (compared with only 53\% when $c$ = 3), and
45\% of the stars in the population shown by the shaded histogram
(dust) experienced flux values greater than the benchmark value
(compared with 33\% when $c$ = 3).

As a final point of analysis, we consider what role our assumed
distribution of $q$ values (which determine orbital angular momentum)
has on the results presented above. Toward that end, we recalculate
the flux distribution shown in Figure 11, but use purely circular
orbits (i.e., we set $q = q_{max}$). The resulting flux distributions
are shown in Figure 14. Not surprisingly, this change has relatively
little effect on the flux experienced by stellar members for the case
with no dust, where 46\% of stars experience flux values greater than
the benchmark value (compared with 53\% found in Figure 11).  However,
adopting circular orbits means that stars born outside of the central
void will always remain outside. As a result, the median of the
distribution representing the case including dust attenuation is
significantly lower for the population shown in Figure 14 than for
that shown in Figure 11. In Figure 14, only about 15\% of the stars in
the population shown by the shaded histogram (with dust) experience
FUV flux values greater than our benchmark value, or about half of the
corresponding value shown in Figure 11. 

\section{Conclusion}

This paper explores the effects of background FUV radiation fields on
the planet forming potential of young stars within intermediate-sized
clusters. Our results are summarized as follows: 

We have determined the distributions of FUV flux levels in young
clusters, including the effects of stellar orbits.  The orbits of
stars in clusters can be described in terms of two dimensionless
orbital parameters, the dimensionless energy $\epsilon$ and angular
momentum $q$.  A star's orbital period depends primarily on the energy
and is longer for larger clusters.  Despite a large range in their
periods, almost all stars born in intermediate sized clusters complete
at least one orbit before stellar outflows, radiation, and supernovae
remove the surrounding gas and dust (which is expected to take $\sim 3
- 5$ Myrs), and typically execute several orbits before planets form
within their circumstellar disks ($\sim 10$ Myr). As a result,
orbit-averaged flux values provide useful measures from which to gauge
how efficiently the cluster background FUV radiation evaporates disks,
and thereby limits planet formation. In this paper, we have
constructed distributions of the FUV flux levels that are expected in
young embedded clusters (see Figures 10 -- 14).  These distributions
are calculated both with and without the effects of dust attenuation,
where the former provides a lower bound and the latter provides an
upper bound to the expected radiation levels.

This paper also calculates the values of the orbit-averaged FUV flux
impinging upon star/disk systems as a function of orbital energy
$\epsilon$ (see Figures 7 -- 9). These calculations are carried out
for three values of orbital angular momentum $q/q_{max}$ and for 27
cluster profiles, defined through a choice of stellar membership $N$,
star formation efficiency $\eta$, density profile index $c$, and an
assumed dust scenario. Although a full description of cluster
radiation fields requires the full distribution of fluxes as shown in
Figures 10 -- 14, we can summarize the results by presenting the
values of dimensionless energy $\epsilon$ = $\epsilon_b$ for which the
orbit-averaged flux equals our adopted benchmark value $G_B$ = 3000
$G_0$ (see Table 2). [Note that this value of flux causes disks
associated with one solar mass stars to experience disruption in giant
planet formation.] In general, the orbit-averaged flux experienced by
a star/disk system in a given cluster environment increases with the
orbital energy $\epsilon$ and decreases with the orbital angular
momentum $q/q_{max}$. For similar orbital parameters, stars experience
higher flux values when in more populated clusters, clusters with
steeper density profiles, and in clusters with greater star formation
efficiencies.

Given the steep nature of the IMF, and the relatively modest
membership sizes $N$ considered herein, the efficiency with which
circumstellar disks are evaporated by background FUV fields varies
greatly from cluster to cluster.  Descriptions of FUV radiation fields
must thus be represented in a statistical manner. Here we calculate
the probability distribution of the orbit-averaged flux for five
cluster profiles, both with and without dust attenuation. These
results indicate that, in the absence of dust attenuation, between 1/3
and 2/3 of stars in intermediate-sized clusters experience FUV flux
values capable of inhibiting giant planet formation, where this
fraction increases as the cluster size increases from $N$ = 100 to
1000.  Interestingly, although larger clusters are better able to
shield stellar systems via dust attenuation of the central FUV
radiation, roughly the same percentage of stars enter the central void
associated with the most massive star in the system. As a result,
roughly 1/3 of stars will experience FUV flux values greater than our
benchmark value regardless of cluster size.

To summarize, this study shows that a substantial fraction (1/3 to
1/2, depending on dust attenuation) of the solar sytems born within
young embedded clusters are exposed to FUV radiation fields more
intense than the benchmark value $G_B$ = 3000 $G_0$ = 4.8 erg
sec$^{-1}$ cm$^{-2}$. By driving photoevaporation, radiation at this
level will truncate circumstellar disks surrounding solar type stars
to radii $r \sim 36$ AU over 10 Myr; this timescale is comparable to
embedded cluster lifetimes, disk lifetimes, and the time required for
giant planets to form. The disks around smaller stars, which are more
common, are more easily evaporated. As a result, a large fraction of
solar systems are affected by the background radiation in their birth
clusters. However, the effects are relatively modest: Although the
outer parts of the disks can be truncated, regions near $r \sim 5$ AU,
where planets are most easily formed, are generally left unscathed. As
a result, these radiative effects are neither negligible nor dominant.
In addition, the FUV radiation fields produced within young clusters
is characterized by a wide distribution, so that assessments of
radiative effects must be made statistically.

An important challenge for the future is to compare the theoretical
predictions of this work with observations. Since photoevaporation
acts to disperse circumstellar material, disk lifetimes should vary
with cluster environment. As outlined above, however, one expects a
wide range of radiative fluxes --- and hence evaporation rates and
disk lifetimes --- within the same cluster. One must thus compare the
full distribution of disk lifetimes for a given cluster with the
corresponding distributions in other clusters.  This comparison
requires a great deal of data, which should be forthcoming in the next
decade.  In addition to disk lifetimes, one can also use planets as a
diagnostic: A number of stars in the solar neighborhood are reported
to have planetary companions that can be directly imaged, including
HR8799 (Marois et al. 2008), Formalhaut (Kalas et al. 2008), and Beta
Pic (Lagrange et al. 2009).  These imaged planets reside (and
presumably formed) in the outer parts of their solar systems, where
circumstellar disks are most easily evaporated.  With expected
improvements in observational capability, future searches for
planetary objects at wide separations (where they can be imaged) will
be carried out. The results of this paper suggest that the success
rate of these searches will be a function of the star formation
environment: Intense UV radiation fields should lead to a suppression
of observed planets with large semimajor axes (roughly $a > 40$ AU). 

\acknowledgments

We are grateful to Marco Fatuzzo for many useful discussions.  Lisa Holden, Edward
Landis and Jeremy Spitzig were supported at NKU through a 2009 CINSAM
Research Grant.  Edward Landis was also supported through a grant from
the Kentucky Space Grant Consortium (grant 4000517KSGC).  FCA is
supported by the University of Michigan through the Michigan Center
for Theoretical Physics, by NASA through the Origins of Solar Systems
Program (grant NNX07AP17G), by NSF through the Division of Applied
Mathematics (grant DMS-0806756), and by the Foundational Questions
Institute (grant RFP1-06-1).

\clearpage

\begin{deluxetable}{rccccccc}
\tablecolumns{8}
\tablewidth{0pc}
\tablecaption{Results from Cluster Simulations}
\tablehead{
\colhead{}    &  \multicolumn{3}{c}{Mean Values} &   \colhead{}   &
\multicolumn{3}{c}{Median Values} \\
\cline{2-4} \cline{6-8} \\
\colhead{N} & \colhead{$m_{max}$}   & \colhead{$L_{FUV}^{max}/\langle L_{FUV}^*\rangle$}    
& \colhead{$L_{FUV}^c/\langle L_{FUV}^*\rangle$} &
\colhead{}    & \colhead{$m_{max}$}   & \colhead{$L_{FUV}^{max}/\langle L_{FUV}^*\rangle$}    
& \colhead{$L_{FUV}^c/\langle L_{FUV}^*\rangle$}}
\startdata
100 & 13 &90 &100 &
& 8.1 & 6.4 &8.3 \\
300 & 24 &230 &300 &
& 17 & 72 &110 \\
1000 & 42 &540 &1000 &
& 36 & 390 &760 \\
\enddata
\end{deluxetable}
\clearpage

\begin{deluxetable}{rccccccc}
\tablecolumns{8}
\tablewidth{0pc}
\tablecaption{Results from Orbit Averaged Flux Calculations}
\tablehead{
 \multicolumn{4}{c}{Cluster Parameters} &   \colhead{}   &
\multicolumn{3}{c}{$\epsilon_b$ vs. $q/q_{max}$} \\
\cline{1-4} \cline{6-8} \\
\colhead{$N$}    & \colhead{$\eta$} &   \colhead{$c$}   &
\colhead{$Q$} & \colhead{} &\colhead{0.2} & \colhead{0.5} & \colhead{0.8}}
\startdata
100 & 1/3 &3 &no dust & & 0.68  & 0.73 & 0.75\\
100 & 1/3 &3 &10& & 0.68 & 0.73  & 0.75 \\
100 & 1/3 &3 &3 & & 0.70 & 0.74  & 0.75 \\

100 & 1/3 &4 &no dust & & 0.58 & 0.63  & 0.66 \\
100 & 1/3 &4 &10 & & 0.58 & 0.63 & 0.66  \\
100 & 1/3 &4 & 3& & 0.60 & 0.66  & 0.68 \\

100 & 1/5 &3 &no dust & & 0.68  & 0.73 & 0.75\\
100 & 1/5 &3 &10& & 0.69 & 0.73  &  0.75\\
100 & 1/5 &3 &3 & & 0.71 & 0.76 & 0.78 \\

300 & 1/3 &3 &no dust & & 0.52  & 0.58 & 0.60\\
300 & 1/3 &3 &10& &0.55  & 0.60  &  0.62\\
300 & 1/3 &3 &3 & &0.58  & 0.66  & 0.70 \\

300 & 1/3 &4 &no dust & & 0.40 & 0.46  & 0.48 \\
300 & 1/3 &4 &10& & 0.43 & 0.49  & 0.51 \\
300 & 1/3 &4 &3 & & 0.48 & 0.57 & 0.63  \\

300 & 1/5 &3 &no dust & & 0.52  & 0.58 & 0.60\\
300 & 1/5 &3 &10& & 0.56 & 0.63  & 0.66 \\
300 & 1/5 &3 &3 & &0.61  & 0.70  & 0.76  \\

1000 & 1/3 &3 &no dust & & 0.46  & 0.51 & 0.54\\
1000 & 1/3 &3 &10& & 0.51 & 0.58  & 0.61 \\
1000 & 1/3 &3 &3 & & 0.57 & 0.67  & 0.74 \\

1000 & 1/3 &4 &no dust & & 0.33 & 0.39  & 0.41 \\
1000 & 1/3 &4 &10 & &  0.39& 0.47 & 0.53  \\
1000 & 1/3 &4 &3& &0.47  & 0.60  &0.68  \\

1000 & 1/5 &3 &no dust & & 0.46  & 0.51 & 0.54\\
1000 & 1/5 &3 &10& & 0.53 & 0.62  & 0.67 \\
1000 & 1/5 &3 &3 & & 0.61 & 0.73  & 0.79 \\

\enddata
\end{deluxetable}


\begin{figure}
\epsscale{1.0}
\plotone{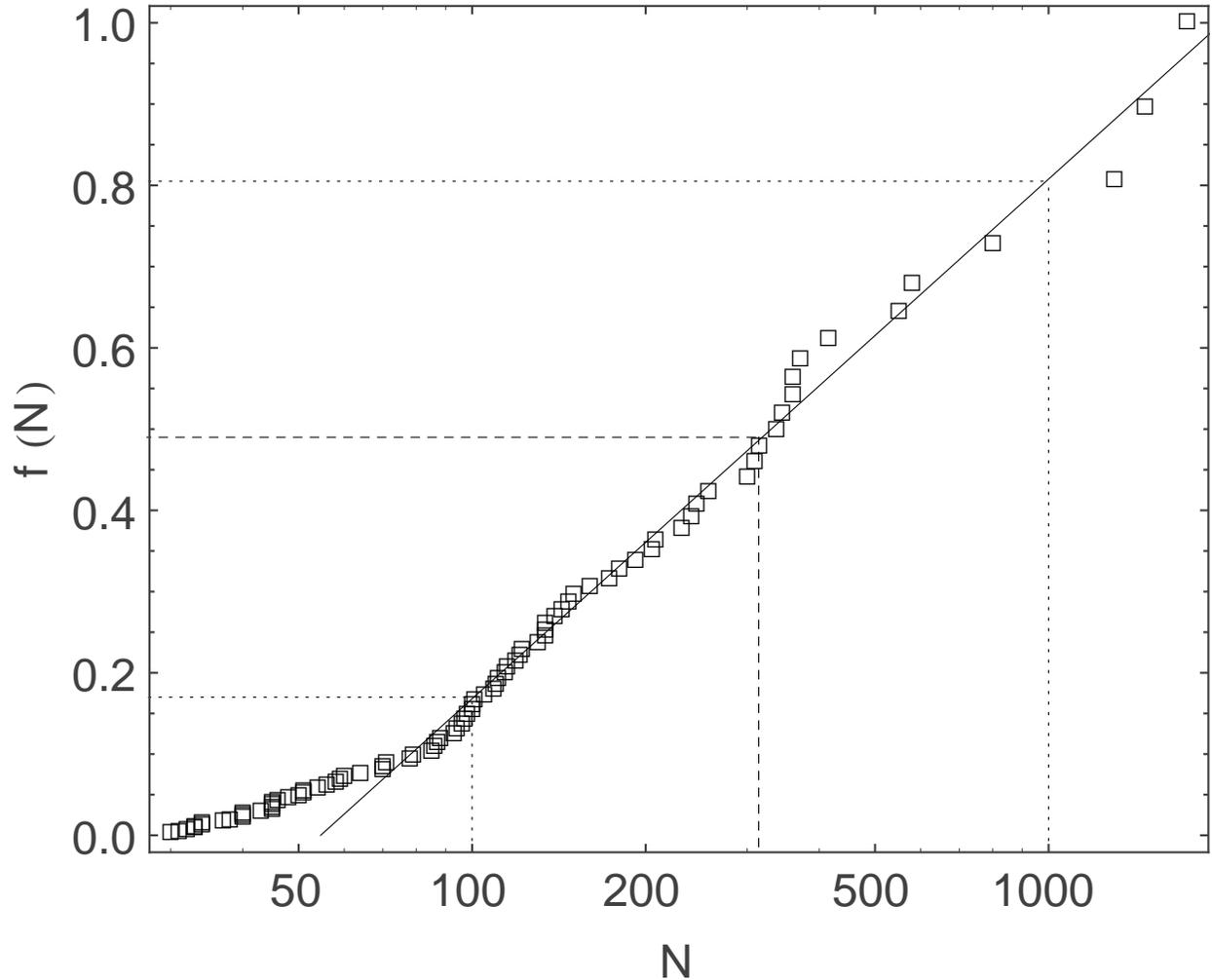}
\caption{Cumulative distribution showing the fraction of stars born in
clusters of stellar membership $N$ as a function of $N$.  The data
points show the combined observed distribution as compiled in Lada \&
Lada (2003) and Porras et al. (2003).  The solid line serves as a
reference to illustrate that the number of stars born in clusters of
size $N$ is evenly distributed logarithmically for intermediate sized
clusters.  The dotted and dashed lines mark the references values 
discussed in the text. \label{fig1}}
\end{figure}

\clearpage

\begin{figure}
\epsscale{1.0}
\plotone{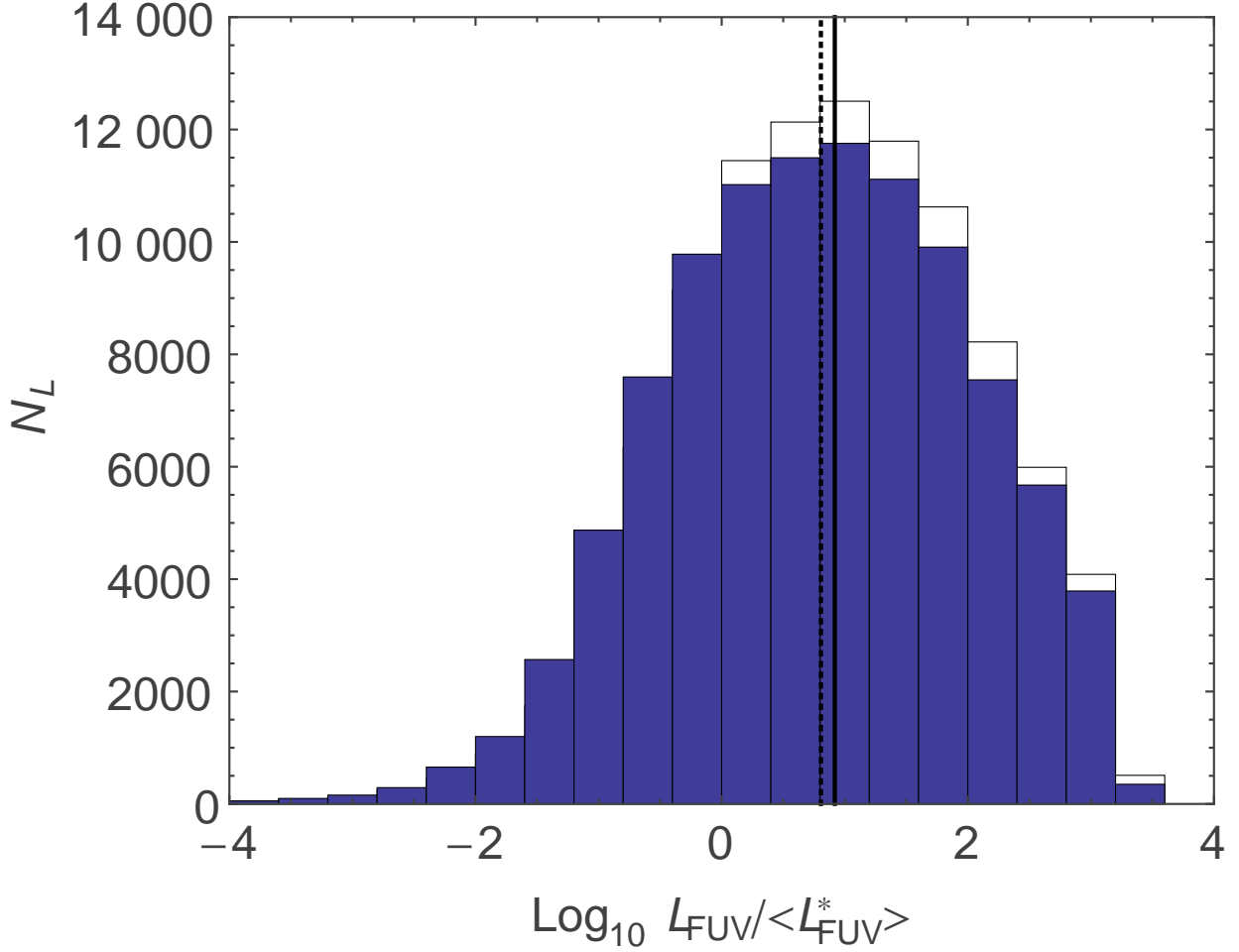}
\caption{Histogram of Log$_{10}\,[ L_{FUV}^{max}/\langle L_{FUV}^*
\rangle]$ (shaded) and Log$_{10}\,[ L_{FUV}^{c}/\langle L_{FUV}^*
\rangle]$(unshaded) for $N = 100$.  The distribution was obtained by
running 100,000 simulations in which we sample the IMF 100 times to
specify the masses and luminosities of the 100 stars that populate
each theoretical cluster. The vertical solid and dashed lines denote
the median values of luminosity for the cluster and for the most
massive star, respectively. We note that the mean value of
$L_{FUV}^{max}/\langle L_{FUV}^*\rangle$ is 90 and the mean value of
$L_{FUV}^c/\langle L_{FUV}^*\rangle$ is 100.\label{fig2}}
\end{figure}

\clearpage

\begin{figure}
\epsscale{1.0}
\plotone{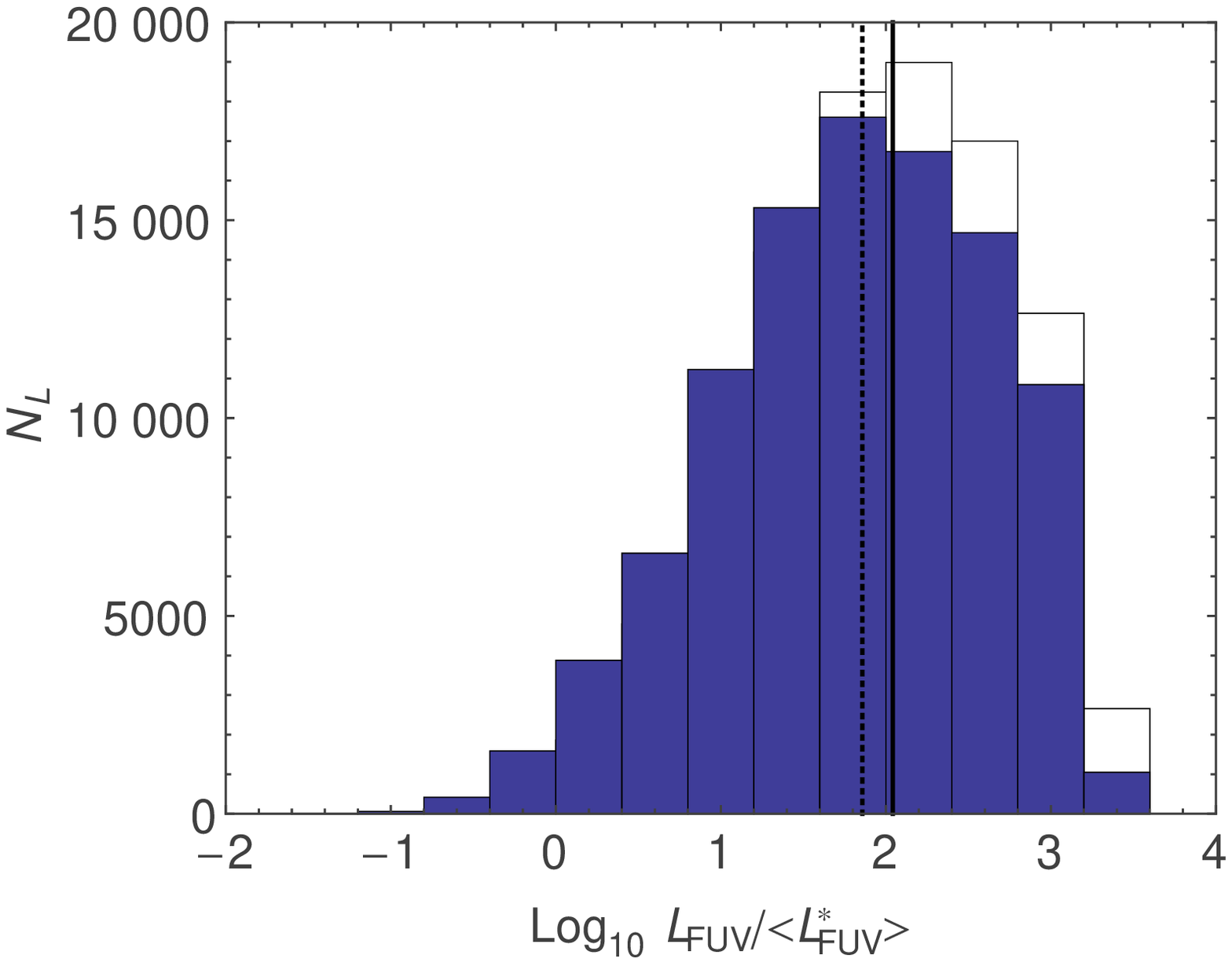}
\caption{Same as Figure 2, but with $N = 300$. 
We note that the mean value of $L_{FUV}^{max}/\langle
L_{FUV}^*\rangle$ is 230 and the mean value of $L_{FUV}^c/\langle
L_{FUV}^*\rangle$ is 300.\label{fig3}}
\end{figure}

\clearpage

\begin{figure}
\epsscale{1.0}
\plotone{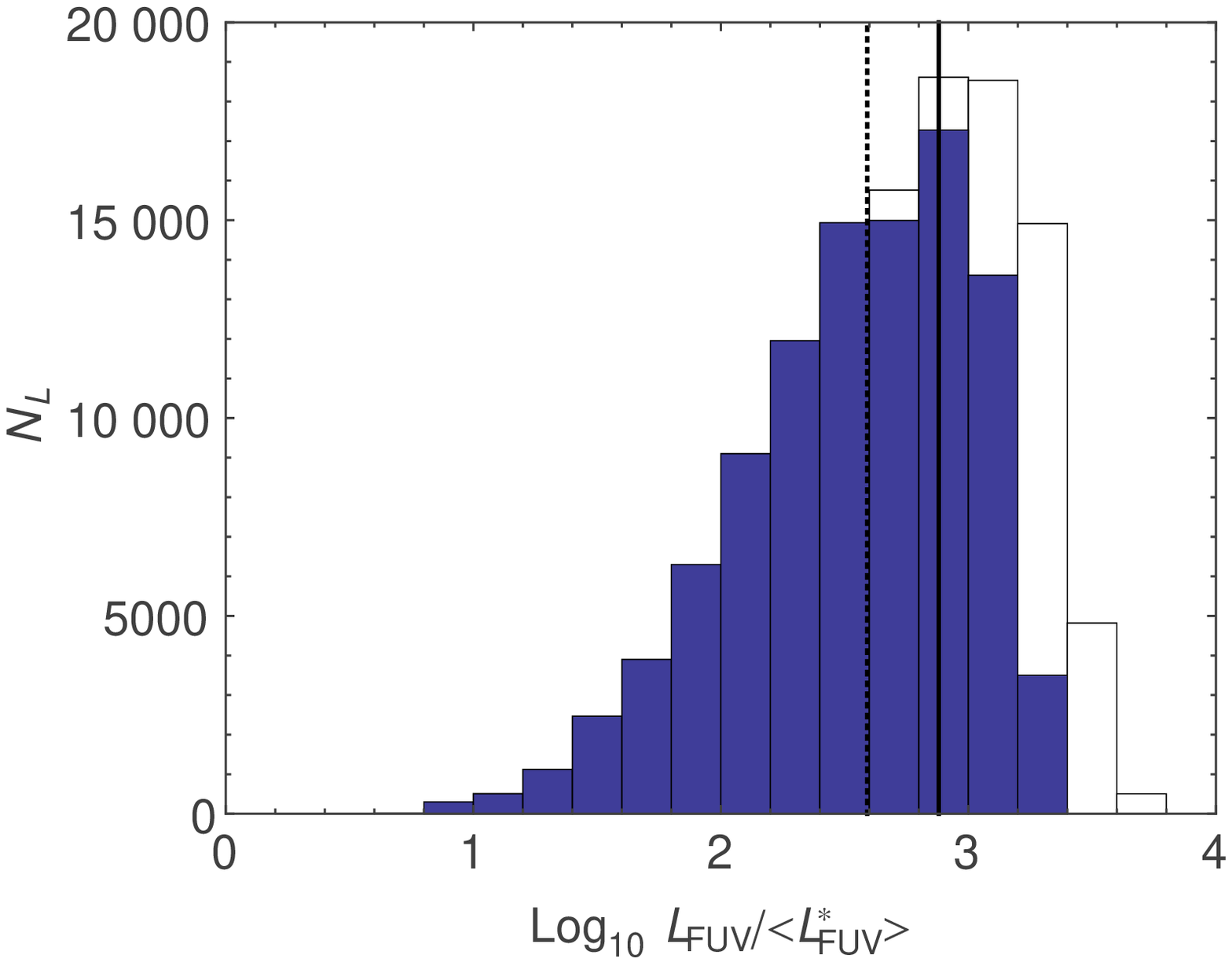}
\caption{Same as Figure 2, but with $N = 1000$. 
We note that the mean value of $L_{FUV}^{max}/\langle
L_{FUV}^*\rangle$ is 540 and the mean value of $L_{FUV}^c/\langle
L_{FUV}^*\rangle$ is 1000.\label{fig4}}
\end{figure}

\clearpage

\begin{figure}
\epsscale{1.0}
\plotone{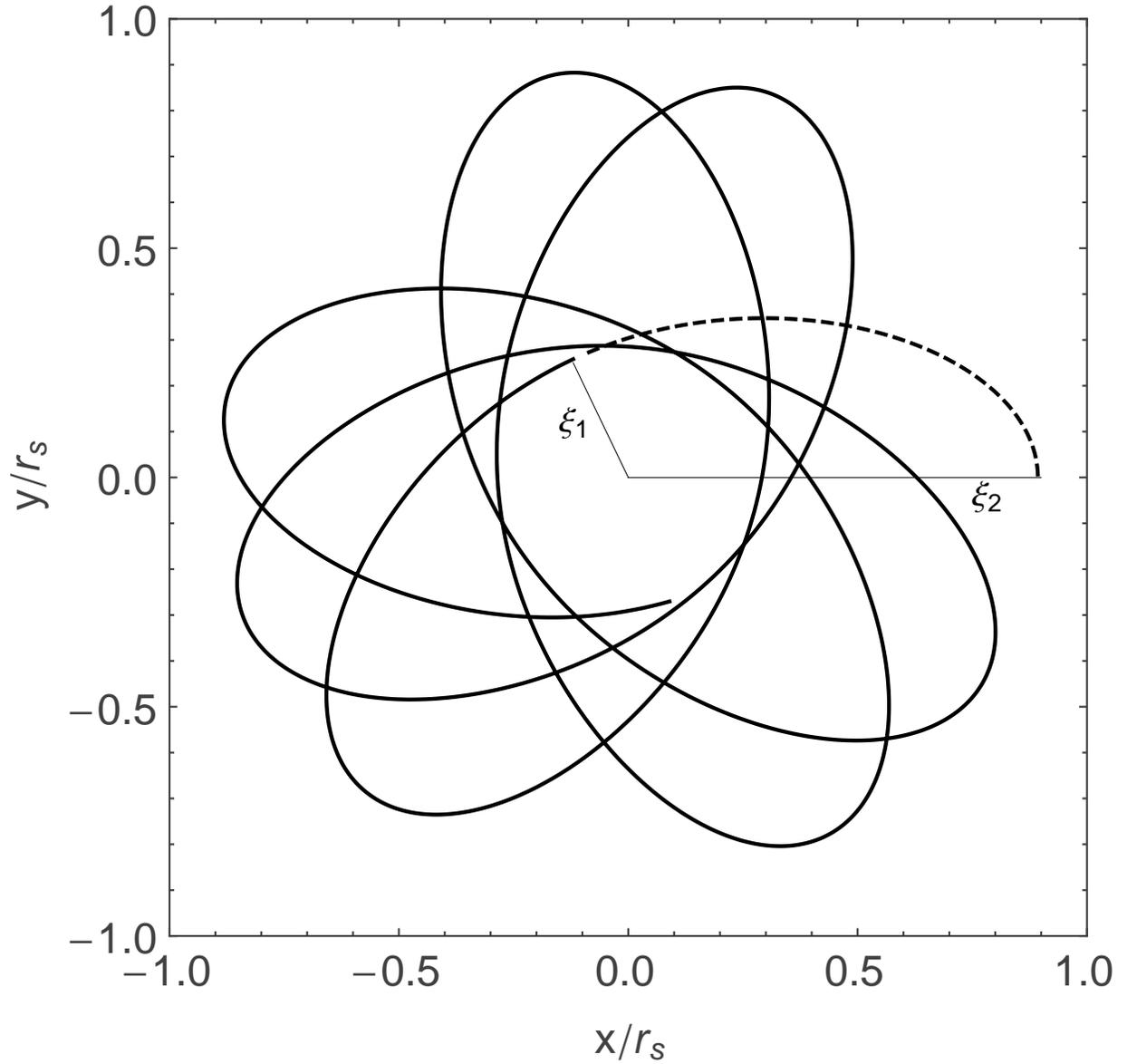}
\caption{Spirographic orbit for dimensionless energy $\epsilon = 0.5$, 
angular momentum variable $q/q_{max} = 0.5$, stellar membership $N$ =
100, and density profile index $c = 3$ (see text). The dashed curve
depicts a half-orbit with the turning points $\xi_1$ = 0.28 and 
$\xi_2$ = 0.89. \label{fig5}}
\end{figure}

\clearpage

\begin{figure}
\epsscale{1.0}
\plotone{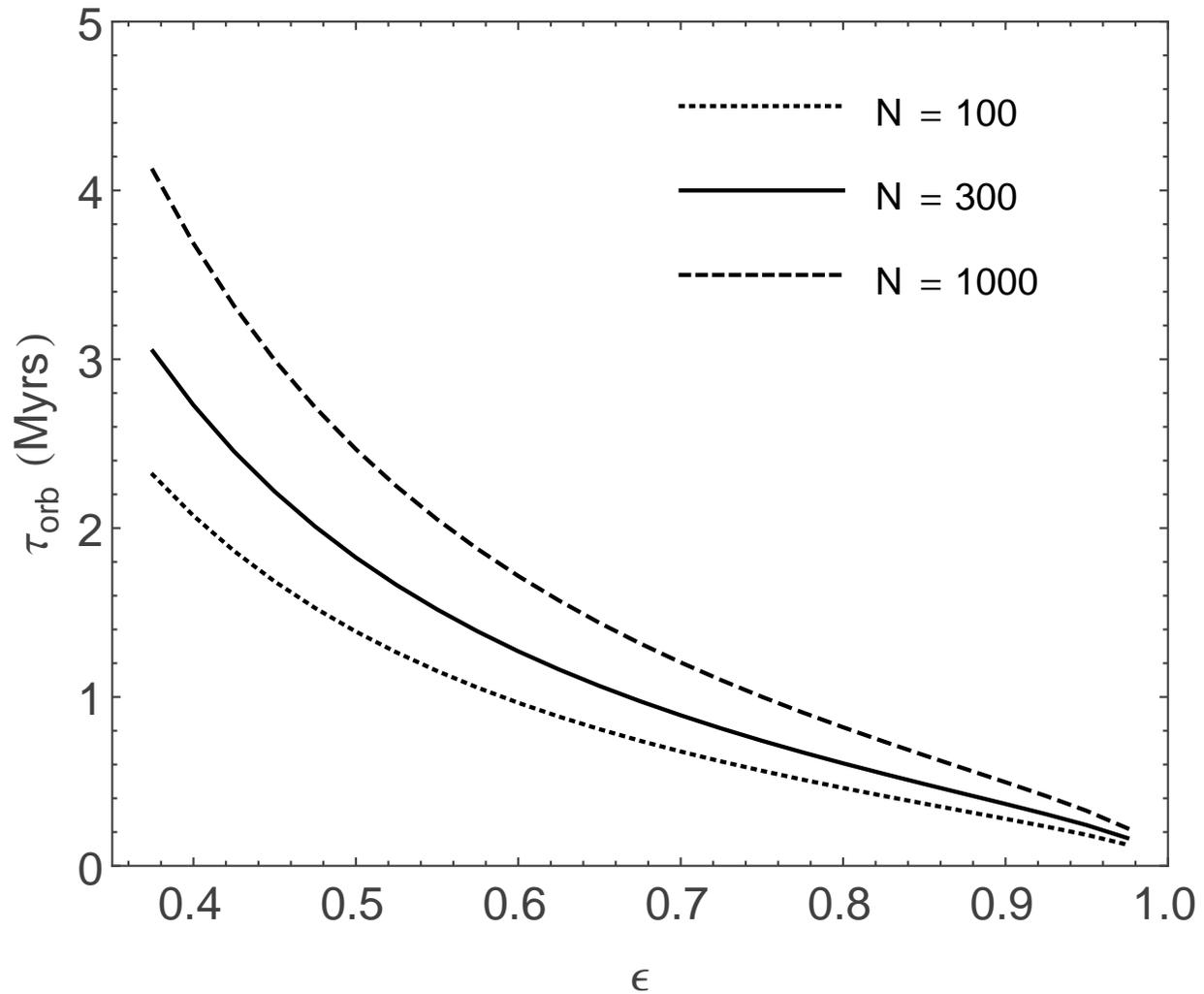}
\caption{Orbital period as a function of dimensionless energy $\epsilon$  
for stars in clusters with membership $N = 100$ (dotted curve), $300$
(solid curve), and $1000$ (dashed curve). For each case, the clusters 
have Hernquist density profiles ($c = 3$) and star formation efficiencies
$\eta = 1/3$. \label{fig6}}
\end{figure}

\clearpage

\begin{figure}
\epsscale{1.0}
\plotone{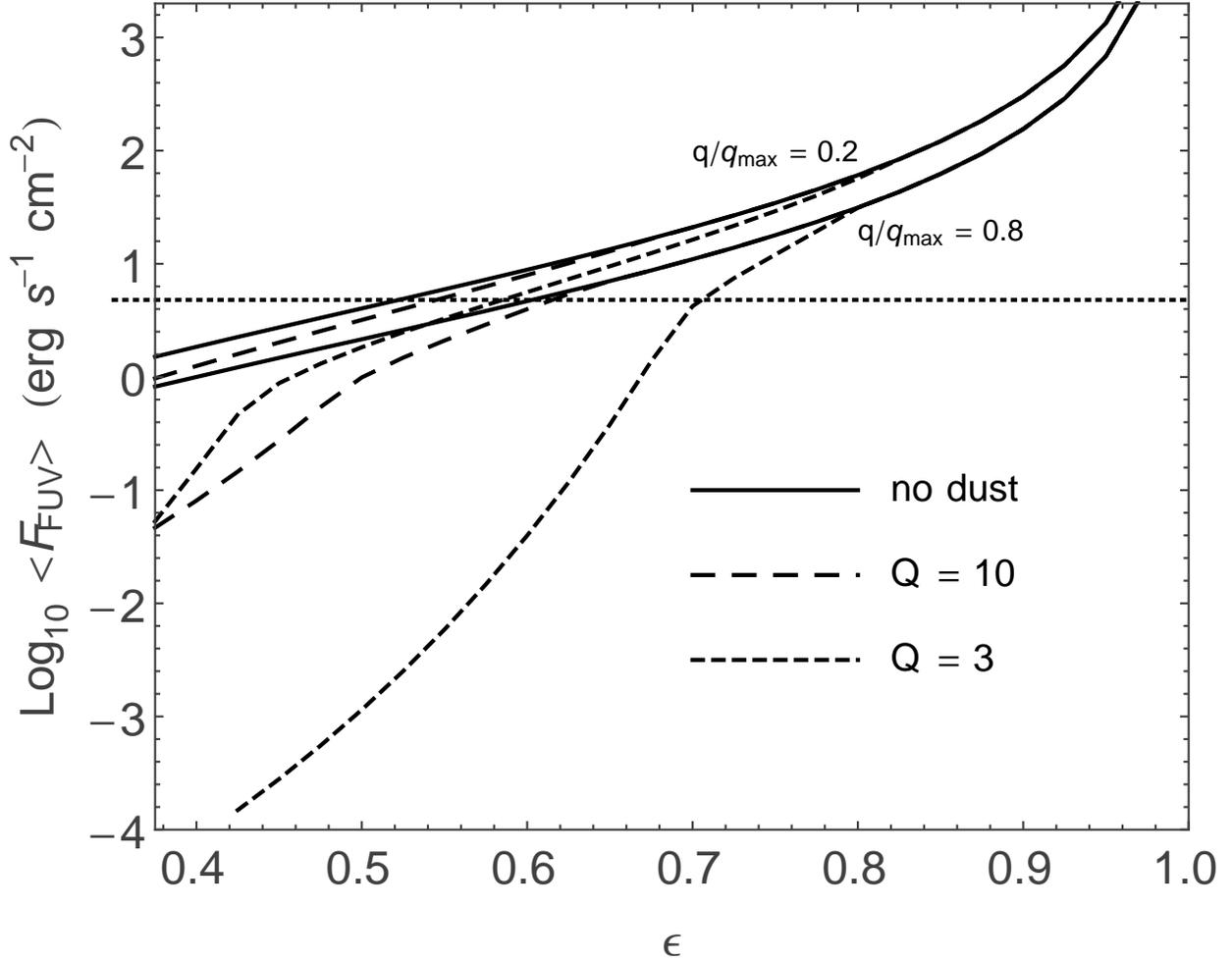}
\caption{Orbit-averaged flux as a function of dimensionless energy
$\epsilon$ for $q/q_{max} = 0.2$ (upper branch) and $q/q_{max}$ = 0.8
(lower branch) and three dust-extinction scenarios: no dust (solid  
curve); $Q = 10$ (long-dashed curve); and $Q = 3$ (short-dashed
curve). For each case, clusters have Hernquist density profiles 
($c = 3$), star formation efficiencies $\eta = 1/3$, and $N = 300$
stellar members.  The dotted line represents the benchmark value as
described in the text.\label{fig7}}
\end{figure}

\clearpage

\begin{figure}
\epsscale{1.0}
\plotone{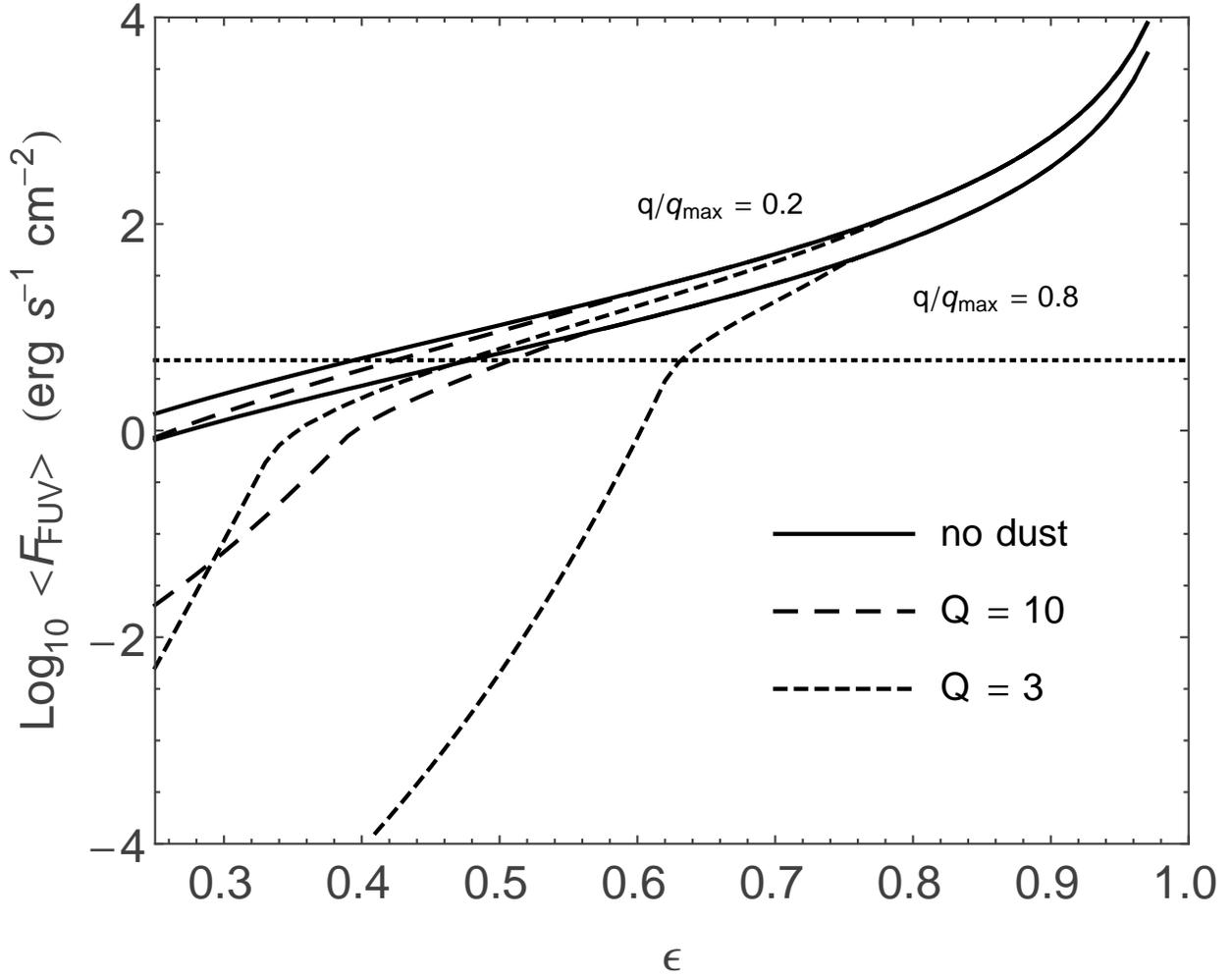}
\caption{Same as Figure 7, but for clusters with modified 
Hernquist density profiles ($c = 4$).
\label{fig8}}
\end{figure}

\clearpage

\begin{figure}
\epsscale{1.0}
\plotone{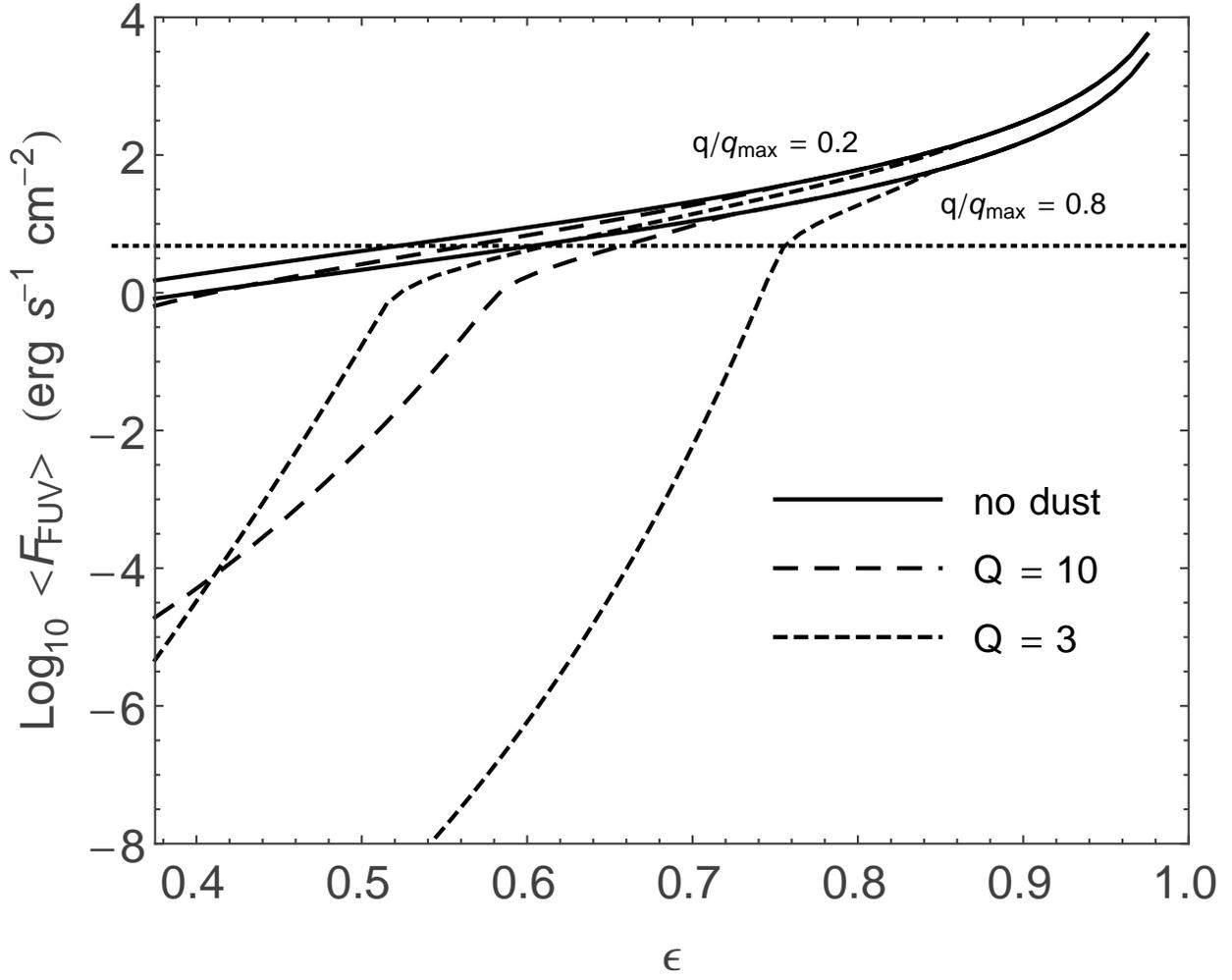}
\caption{Same as Figure 7, but for clusters with star 
formation efficiencies $\eta = 1/5$.\label{fig9}}
\end{figure}

\clearpage

\begin{figure}
\epsscale{1.0}
\plotone{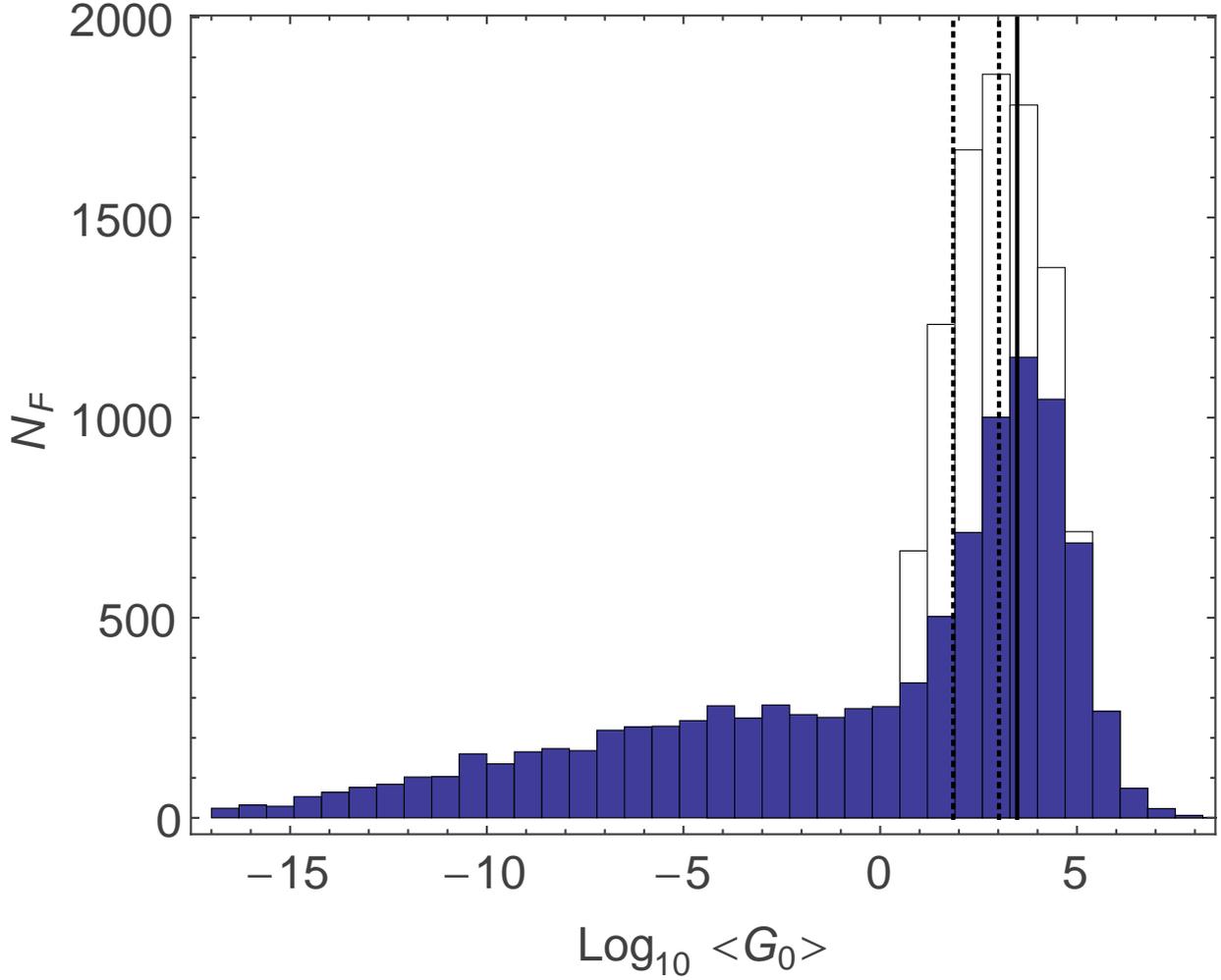}
\caption{Orbit-averaged flux distributions for a sample of 10,000 stars
with randomly selected orbital parameters $\epsilon$ and $q/q_{max}$ (as
described in the text) for a cluster environment with $N = 100$, $c = 3$, 
and $\eta = 1/5$.  Flux values are expressed here in units of $G_0$ (where
$G_0 = 1$ corresponds to a flux value of $1.6\times 10^{-3}$ ergs
s$^{-1}$ cm$^{-2}$, which is close to the value of the interstellar
FUV radiation field).  The unshaded histogram represents the case with no
dust.  The shaded histogram represents the case with dust attenuation
and $Q = 3$.  The solid line represents our benchmark value, as
described in the text. The dotted lines represent the median values of
the two distributions, with the lower value corresponding to the case
that includes dust attenuation.\label{fig10}} 
\end{figure}

\clearpage

\begin{figure}
\epsscale{1.0}
\plotone{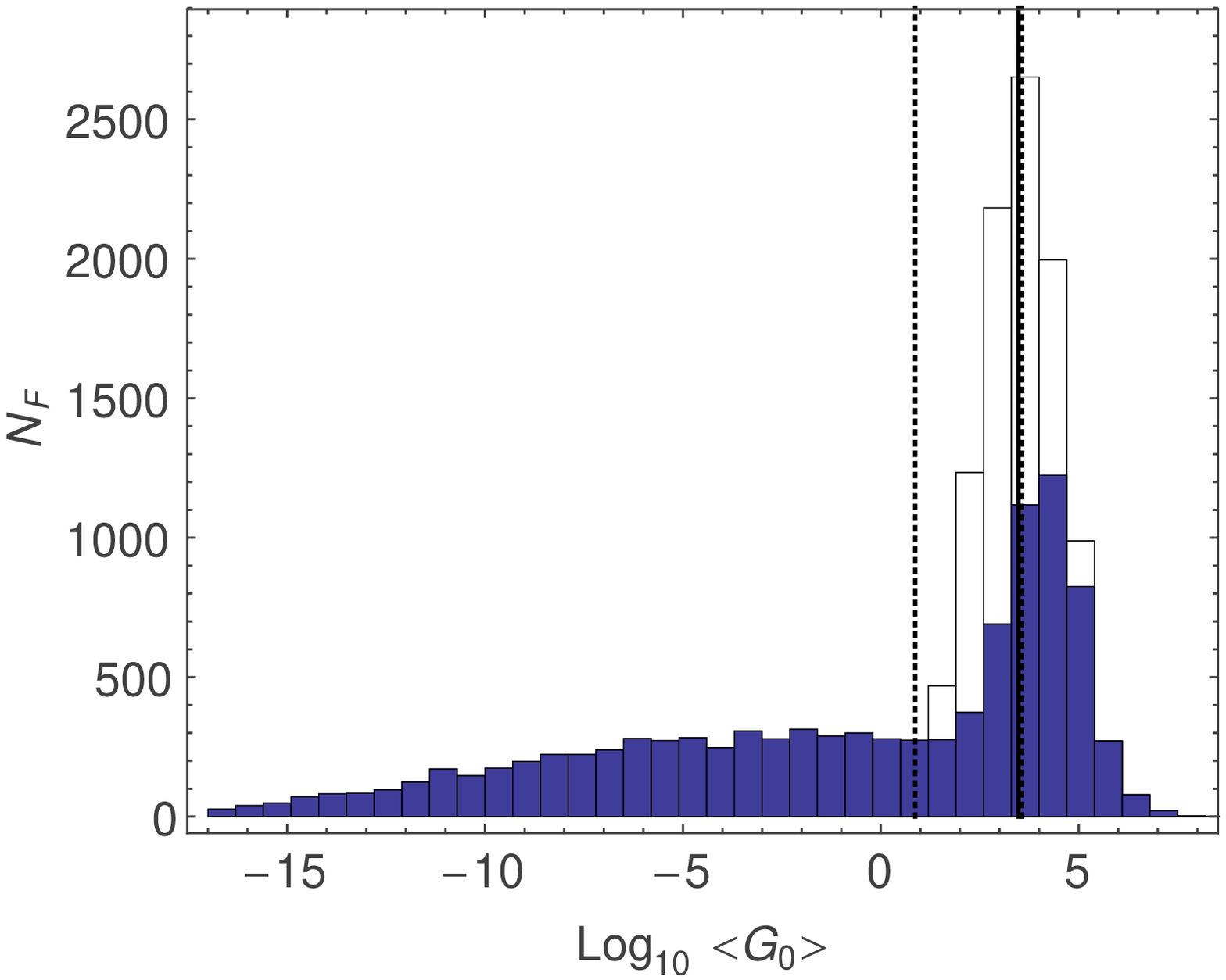}
\caption{Same as Figure 10, but with stellar membership $N = 300$.\label{fig11}}
\end{figure}

\clearpage

\begin{figure}
\epsscale{1.0}
\plotone{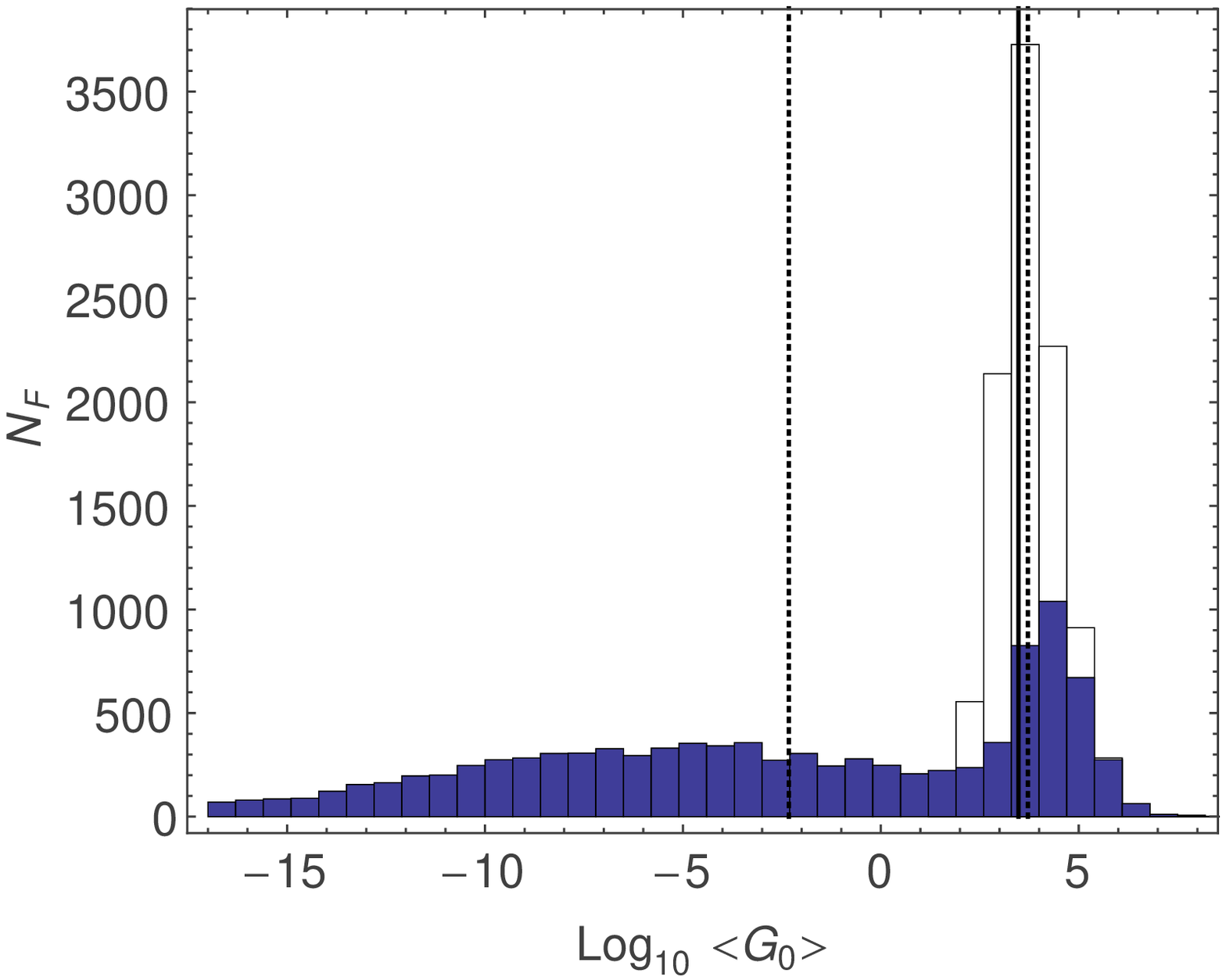}
\caption{Same as Figure 10, but with stellar membership $N = 1000$.\label{fig12}}
\end{figure}

\clearpage

\begin{figure}
\epsscale{1.0}
\plotone{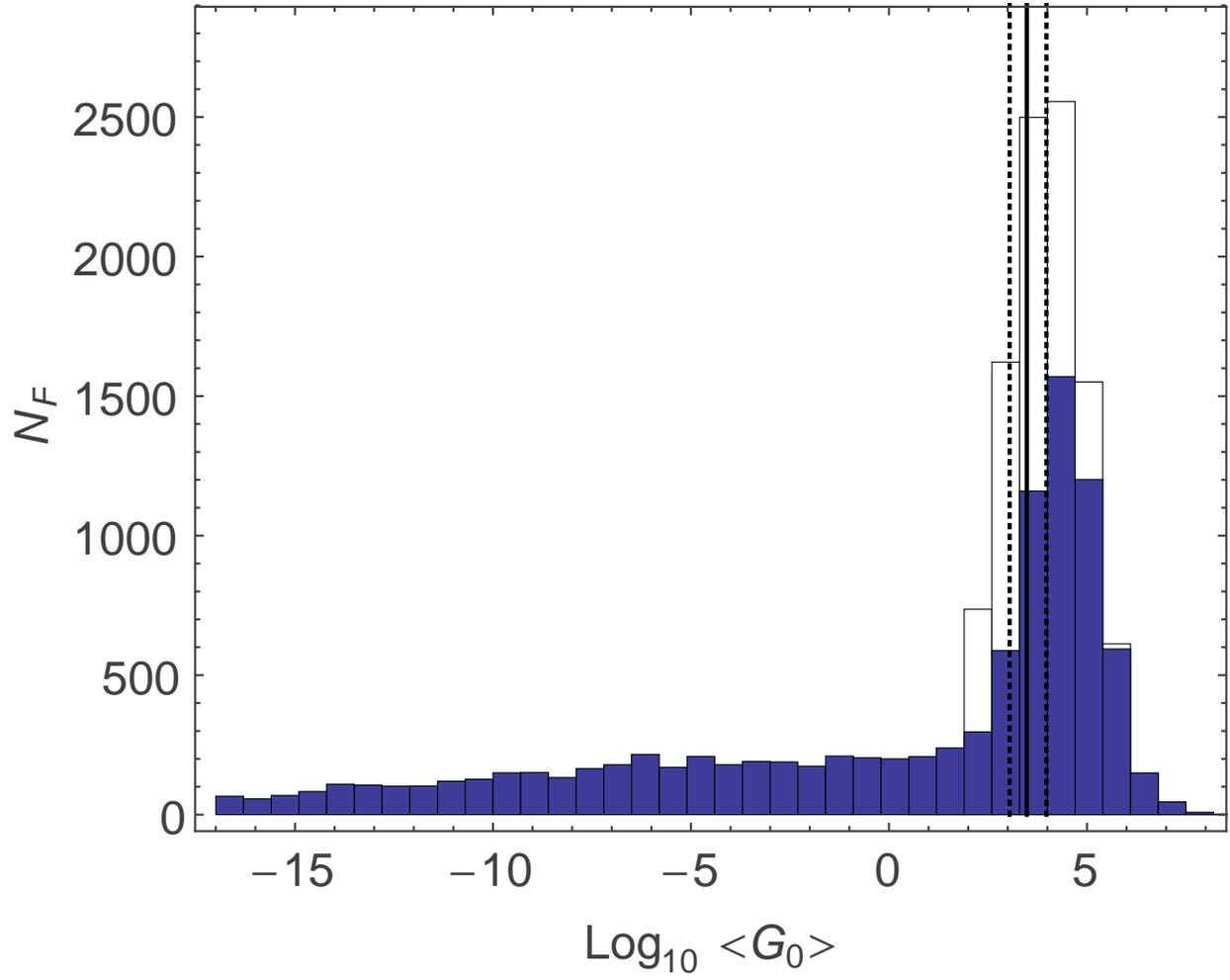}
\caption{Same as Figure 11 ($N = 300$, $\eta = 1/5$, $Q=3$), 
but with density profile index $c=4$.\label{fig13}}
\end{figure}

\clearpage

\begin{figure}
\epsscale{1.0}
\plotone{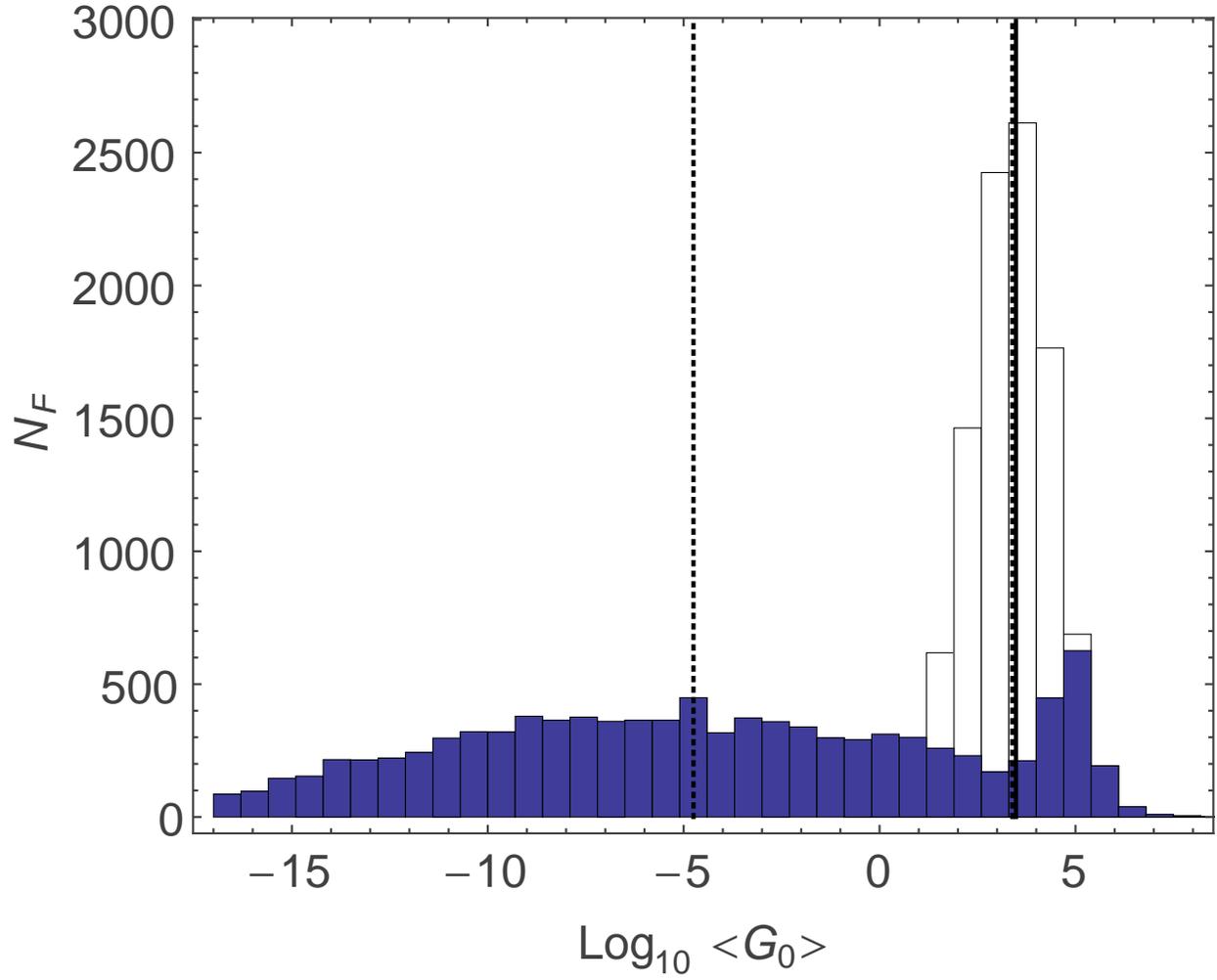}
\caption{Same as Figure 11 ($N = 300$, $\eta = 1/5$, $Q=3$, $c=3$), 
but for circular orbits ($q = q_{max}$).\label{fig14}}
\end{figure}

\clearpage

\end{document}